\documentclass[11pt]{article}
\usepackage{a4p}
\usepackage{amssymb}
\usepackage{epsfig}
\usepackage{cite}
\usepackage{rotating}
\newcommand{\epem}{\ensuremath{\mathrm{e}^+\mathrm{e}^-}}

\newcommand{\tptm}{\ensuremath{\tau^+\tau^-}}
\newcommand{\lplm}{\ensuremath{\ell^+\ell^-}}
\newcommand{\Zz}{\ensuremath{{\mathrm{Z}^0}}}

\newcommand{\WW}{\ensuremath{\mathrm{W}^+\mathrm{W}^-}}
\newcommand{\Wp}{\ensuremath{\mathrm{W}^+}}

\newcommand{\Wm}{\ensuremath{\mathrm{W}^-}}
\newcommand{\eeWW}{\ensuremath{\epem\rightarrow\WW}}
\newcommand{\qq}{\ensuremath{\mathrm{q\overline{q}}}}
\newcommand{\QQ}{\ensuremath{\mathrm{q\overline{q}}}}

\newcommand{\lnu}{\ensuremath{\ell\overline{\nu}_{\ell}}}

\newcommand{\WWqqlv}{\ensuremath{\WW\rightarrow\qq\lnu}}
\newcommand{\WWqqqq}{\ensuremath{\WW\rightarrow\qq\qq}}

\newcommand{\ZZqqqq}{\ensuremath{(\Zz/\gamma)(\Zz/\gamma)\rightarrow\QQ\QQ}}

\newcommand{\eeqq}{\ensuremath{\ee\rightarrow\QQ}}
\newcommand{\eell}{\ensuremath{\ee\rightarrow\lplm}}

\newcommand{\eeqqg}{\ensuremath{\epem\rightarrow \mathrm{q\overline{q}g}}}
\newcommand{\Zqq}{\ensuremath{\Zz/\gamma\rightarrow\QQ}}
\newcommand{\Ztt}{\ensuremath{\Zz/\gamma\rightarrow\tptm}}

\newcommand{\Mw}{\ensuremath{M_{\mathrm{W}}}}
\newcommand{\Gw}{\ensuremath{\Gamma_{\mathrm{W}}}}

\newcommand{\Opal}{OPAL}

\newcommand{\Delphi}{DELPHI}

\newcommand{\Jetset}{JETSET}

\newcommand{\Kandy}{KandY}
\newcommand{\Koralw}{KORALW}
\newcommand{\KKff}{\mbox{KK2f}}

\newcommand{\grcff}{\mbox{grc4f}}
\newcommand{\Pythia}{PYTHIA}

\newcommand{\Ariadne}{ARIADNE}

\newcommand{\PHOJET}{PHOJET}

\newcommand{\Herwig}{HERWIG}

\newcommand{\GeV}{\ensuremath{\mathrm{GeV}}}
\newcommand{\GeVc}{\ensuremath{\mathrm{GeV/c}}}
\newcommand{\Prec}{\ensuremath{P^{\mathrm{reco}}}}

\newcommand{\Rn}{\ensuremath{R_N}}

\newcommand{\kI}{\ensuremath{k_I}}
\newcommand{\SK}{\mbox{SK}}
\newcommand{\SKI}{\mbox{SK-I}}
\newcommand{\SKII}{\mbox{SK-II}}
\newcommand{\SKIIpr}{\mbox{SK-II$^{\prime}$}}
\newcommand{\ARI}{\mbox{AR-1}}
\newcommand{\ARII}{\mbox{AR-2}}
\newcommand{\ARIII}{\mbox{AR-3}}

\newcommand{\Rflow}{\ensuremath{R_{\mathrm{flow}}}}
\newcommand{\sigRnstat}{\ensuremath{\sigma^{\mathrm{stat.}}_{\Rn}}}
\newcommand{\sigRntot}{\ensuremath{\sigma^{\mathrm{total}}_{\Rn}}}
\newcommand{\sensRn}{\ensuremath{\Delta \Rn/\sigRnstat}}

\newcommand{\roots}{\ensuremath{\sqrt{s}}}

\newcommand{\Ebeam}{\ensuremath{E_{\mathrm{beam}}}}

\newcommand{\Zgamma}{\ensuremath{\Zz/\gamma}}

\newcommand {\ee}         {\ensuremath{\mathrm{e}^+ \mathrm{e}^-}}

\newcommand{\CC}{\mbox{{\sc CC03}}}

\def\etal{\mbox{{\it et al.}}}
\def\ie{\mbox{{\it i.e.}}}
\def\eg{\mbox{{\it e.g.}}}

\def\gappeq{\ensuremath{\mathrel{ \rlap{\raise.5ex\hbox{>}}
                      {\lower.5ex\hbox{\sim}}}}}
\newcommand{\lappeq}{\ensuremath{\mathrel{\rlap{\raise.5ex\hbox{<}}{\lower.5ex\hbox{\sim}}}}}

\newcommand{\PLB}[3]  {Phys.\ Lett.\ \textbf{B#1} (#2) #3}
\newcommand{\ZPC}[3]  {Z.\ Phys.\ \textbf{C#1} (#2) #3}
\newcommand{\EPC}[3]  {Eur.\ Phys.\ J.\ \textbf{C#1} (#2) #3}
\newcommand{\NIMA}[3] {Nucl.\ Instr.\ Meth.\ \textbf{A#1} (#2) #3}

\newcommand{\PRL}[3]  {Phys.\ Rev.\ Lett.\ \textbf{#1} (#2) #3}
\newcommand{\PRD}[3]  {Phys.\ Rev.\ \textbf{D#1} (#2) #3}
\newcommand{\NPB}[3]  {Nucl.\ Phys.\ \textbf{B#1} (#2) #3}

\newcommand{\CPC}[3]  {Comput.\ Phys.\ Commun.\ \textbf{#1} (#2) #3}
\newcommand{\Zzero}{\mbox{${\mathrm{Z}^0}$}}
\newcommand{\ZZ}{\mbox{\Zzero\Zzero}}

\def\opalabbiendi{OPAL Collaboration, G.\ Abbiendi \etal}
\def\opalackerstaff{OPAL Collaboration, K.\ Ackerstaff \etal}
\def\opalalexander{OPAL Collaboration, G.\ Alexander \etal}

\newcommand{\QQQQ}{\ensuremath{\mathrm{4q}}}
\newcommand{\QQLV}{\ensuremath{\mathrm{qq}\ell\nu}}
\newcommand{\nch}{\ensuremath{n_{\mathrm{ch}}}}

\newcommand{\nchQQQQ}{\ensuremath{\langle n_{\mathrm{ch}}^{\QQQQ}\rangle}}
\newcommand{\nchQQLV}{\ensuremath{\langle n_{\mathrm{ch}}^{\QQLV}\rangle}}
\newcommand{\Dnch}{\ensuremath{\Delta\langle n_{\mathrm{ch}}\rangle}}
\newcommand{\nchW}{\ensuremath{\langle n_{\mathrm{ch}}^\mathrm{qq}\rangle}}

\newcommand{\xp}{\ensuremath{x_{p}}}

\begin{document}
%

%
%
\begin{titlepage}
 \begin{center}{\large    EUROPEAN ORGANIZATION FOR NUCLEAR RESEARCH
 }\end{center}\bigskip
\begin{flushright}
  CERN-PH-EP/2005-035 \\  
  15th July 2005 \\
\end{flushright}
\bigskip\bigskip\bigskip\bigskip\bigskip
\begin{center}
 {\huge\bf \boldmath Colour reconnection in \eeWW\ \\
   \vspace{2mm} at $\roots=189$--$209$~\GeV}\\
\end{center}
\bigskip\bigskip
\begin{center}{\LARGE The OPAL Collaboration
}\end{center}\bigskip\bigskip

\bigskip\begin{center}{\large Abstract}\end{center}
\noindent
 The effects of the final state interaction phenomenon known as colour
 reconnection are investigated at centre-of-mass energies in the range
 $\roots\simeq189$--209~\GeV\ using the OPAL detector at LEP\@.
 Colour reconnection is expected to affect observables based on
 charged particles in hadronic decays of \WW.  Measurements of
 inclusive charged particle multiplicities, and of their angular
 distribution with respect to the four jet axes of the events, are
 used to test models of colour reconnection.
  The data are found to exclude extreme scenarios of the
  Sj\"ostrand-Khoze Type I (\SKI) model and are compatible with other
  models, both with and without colour reconnection effects.  In the
  context of the \SKI\ model, the best agreement with data is obtained
  for a reconnection probability of 37\%.  Assuming no colour
  reconnection, the charged particle multiplicity in hadronically
  decaying W bosons is measured to be $\nchW =
  19.38\pm0.05(\mathrm{stat.})\pm0.08 (\mathrm{syst.})$.

\bigskip\bigskip\bigskip\bigskip
\bigskip\bigskip
\begin{center}
(Sbmitted to Eur.\ Phys.\ J.\ C)
\end{center}
\end{titlepage}
\begin{center}{\Large        The OPAL Collaboration
}\end{center}\bigskip
\begin{center}{
G.\thinspace Abbiendi$^{  2}$,
C.\thinspace Ainsley$^{  5}$,
P.F.\thinspace {\AA}kesson$^{  3,  y}$,
G.\thinspace Alexander$^{ 22}$,
G.\thinspace Anagnostou$^{  1}$,
K.J.\thinspace Anderson$^{  9}$,
S.\thinspace Asai$^{ 23}$,
D.\thinspace Axen$^{ 27}$,
I.\thinspace Bailey$^{ 26}$,
E.\thinspace Barberio$^{  8,   p}$
% ...
}\end{center}\bigskip
\bigskip
$^{  1}$School of Physics and Astronomy, University of Birmingham,
Birmingham B15 2TT, UK
\newline
$^{  2}$Dipartimento di Fisica dell' Universit\`a di Bologna and INFN,
I-40126 Bologna, Italy
\newline
$^{  3}$Physikalisches Institut, Universit\"at Bonn,
D-53115 Bonn, Germany
\newline
% ...
$^{  5}$Cavendish Laboratory, Cambridge CB3 0HE, UK
\newline
% ...
$^{  8}$CERN, European Organisation for Nuclear Research,
CH-1211 Geneva 23, Switzerland
\newline
$^{  9}$Enrico Fermi Institute and Department of Physics,
University of Chicago, Chicago IL 60637, USA
\newline
% ...
$^{ 22}$Department of Physics and Astronomy, Tel Aviv University,
Tel Aviv 69978, Israel
\newline
$^{ 23}$International Centre for Elementary Particle Physics and
Department of Physics, University of Tokyo, Tokyo 113-0033, and
Kobe University, Kobe 657-8501, Japan
\newline
% ...
$^{ 26}$University of Victoria, Department of Physics, P O Box 3055,
Victoria BC V8W 3P6, Canada
\newline
$^{ 27}$University of British Columbia, Department of Physics,
Vancouver BC V6T 1Z1, Canada
\newline
% ...
$^{  p}$ now at The University of Melbourne, Victoria, Australia
\newline
% ...
$^{  y}$ now at CERN
\newline
% ...

\newpage


\section{Introduction}
\label{sec:Wprop}
 
 Hadronic data in \epem\ collisions can be characterised by event
 shape distributions and inclusive observables such as charged
 particle multiplicities and momentum spectra.  Measurement of the
 detailed properties of the hadronic sector of \WW\ decays allows the
 question of ``colour reconnection'' (CR) \cite{bib:GPZ} to be
 addressed experimentally, in addition to providing tests of Monte
 Carlo models.  The decay products of the two W boson decays have a
 significant space-time overlap as the separation of their decay
 vertices at LEP2 energies is small compared to characteristic
 hadronisation distance scales.  In the fully hadronic channel this
 may lead to new types of final state interactions.  Colour
 reconnection is the general name applied to the case where such final
 state interactions lead to colour exchange between the decay products
 of the two W bosons.  A modification of the colour flow in this way
 could have a significant influence on the measured mass of the W
 boson, as first noted in \cite{bib:GPZ}.  It is therefore essential
 to ascertain whether or not such effects are present in W decays.  As
 described in \cite{bib:SK}, a precedent is set for such effects in
 colour suppressed B meson decays, \eg\ $\mathrm{B} \rightarrow
 \mathrm{J}/\psi \mathrm{K}$, where there is ``cross-talk'' between
 the two original colour singlets, $\bar{\mathrm c}$+s and
 c+spectator.

  There is general consensus that observable effects of such
 interactions during the perturbative phase are expected to be small
 \cite{bib:SK}.  In contrast, significant interference in the
 hadronisation process is considered to be a real possibility.  With
 the current understanding of non-perturbative QCD, such interference
 can be estimated only in the context of specific models
 \cite{bib:GPZ,bib:SK,bib:PYTHIA,bib:GH,bib:LL,bib:ARIADNE,bib:webbercr,bib:HERWIG,bib:EG1,bib:VNI}.
 Other final state effects such as Bose-Einstein correlations (BEC)
 between identical bosons from different W decays may also influence
 the observed event properties.

 This paper presents two different measurements which are sensitive to
 colour reconnection effects.  The inclusive properties of \WW\ decay
 products have been measured
 \cite{bib:opalmw172,bib:opalcr183,bib:delphi_zpc_1999} and found to
 have limited sensitivity to colour reconnection using the data
 available at LEP2.  Characteristic observables such as the inclusive
 charged particle multiplicity in \WWqqqq\ events, \nchQQQQ, and its
 centre-of-mass energy dependence have been widely used to quantify
 the effect of colour reconnection in \WW\ events
 \cite{bib:GPZ,bib:PYTHIA,bib:SK,bib:GH,bib:LL,bib:ARIADNE,bib:HERWIG,bib:EG1,
 bib:VNI}, and are therefore studied in this paper.  As in
 \cite{bib:opalcr183}, the hadronic part of \WWqqlv\ events is
 compared with \WWqqqq\ events, while the leptonically decaying W is
 excluded.

 More recently, all LEP collaborations have concentrated on studies of
 ``particle flow'' \cite{bib:LEP2YR1,bib:WWOXFORD,bib:pflow1}, a
 generalisation of the well-known ``string effect''
 \cite{bib:string_effect} to the four-jet case of \WWqqqq, as models
 predict \cite{bib:lepewwg} this has a larger sensitivity to colour
 reconnection.  This analysis compares the density of charged
 particles in two regions: the first, between pairs of hadronic jets
 originating from the same W boson, and the second, between pairs of
 hadronic jets which originate from different W bosons.  In the
 absence of colour exchange between the two W bosons, the particle
 density is expected to be larger in the first region.  Colour
 reconnection would lead to a migration of particles into the second
 region, in addition to a change in the total multiplicity.  All data
 in the range 189--209~\GeV\ are studied using the particle flow
 method in this paper, which supersedes previous OPAL analyses on the
 subject \cite{bib:opalmw172,bib:opalcr183}.

 This paper is organised as follows: Section~\ref{sec:selection}
 summarises data and Monte Carlo models used,
 Section~\ref{sec:analysis} describes the inclusive charged particle
 and particle flow analyses and Section~\ref{sec:systematics} the
 estimation of systematic effects.  Sections~\ref{sec:results} and
 \ref{sec:conclusions} discuss the results and draw conclusions.

\section{Data Selection and Monte Carlo Models}
\label{sec:selection}
 This paper uses data corresponding to an integrated luminosity of
 approximately 625~pb$^{-1}$ recorded during 1998--2000 with the OPAL
 detector, which is described fully elsewhere~\cite{bib:detector}.
 The data are separated into samples at six centre-of-mass energies,
 varying between approximately 189~\GeV\ and 209~\GeV.  Those accumulated
 above 202.5~\GeV\ are considered at a single luminosity weighted mean
 centre-of-mass energy of 206.0~\GeV. The selection criteria and
 distribution of data by \WW\ final state, luminosity and
 centre-of-mass energy are given in \cite{bib:opalmw_final}, with a
 total of 5401 \WWqqqq\ and 2757 \WWqqlv\ candidates selected.  Only
 \WWqqlv\ events in which the charged lepton was identified as an
 electron or a muon are used, to ensure that the
 charged particle multiplicity of the hadronically decaying W is well
 understood.

 It is essential to have reliable selection of charged particles in
 the detector in this analysis.  Charged particles may have up to 159
 hits in the principal tracking chamber, the jet chamber. Tracks used
 in the analysis are required to have a minimum of 40 hits in the
 $|\cos\theta|$ region\footnote{The OPAL coordinate system is defined
 such that the origin is at the geometric centre of the jet chamber,
 $z$ is parallel to, and has positive sense along, the e$^-$ beam
 direction, $r$ is the coordinate normal to $z$, $\theta$ is the polar
 angle with respect to +$z$ and $\phi$ is the azimuthal angle around
 $z$.} in which at least 80 are possible.  At larger $|\cos\theta|$,
 the number of hits is required to be at least 50\% of the expected
 number and not fewer than 20, corresponding to a fiducial acceptance
 of $|\cos\theta|<0.96$.  Tracks must have a momentum component in the
 plane perpendicular to the beam axis of greater than 0.15~\GeVc, and
 a measured momentum $p$ of less than 100~\GeV$/c$.  For each track,
 the point of closest approach to the collision axis is found, and the
 distance between this point and the average interaction point is
 required to be less than 2~cm in the \mbox{$r$-$\phi$} plane and less
 than 25~cm in $z$.
  Clusters of energy in the electromagnetic
 calorimeter are required to have a measured energy greater than
 0.10~\GeV\ if they occur in the barrel region of the detector
 ($|\cos\theta|<0.82$), and greater than 0.25~\GeV\ if they occur in
 the endcaps ($0.82<|\cos\theta|<0.98$).

 Most samples of Monte Carlo events used in this paper include
 detailed simulation of the OPAL detector \cite{bib:GOPAL} and of
 initial state photon radiation and have been passed through the same
 selection and analysis procedures applied to the data (``detector
 level'').  A second class of samples does not include initial state
 photon radiation or simulation of the detector and allows all
 particles with lifetimes shorter than $3\times10^{-10}$~s to decay
 (``hadron level'').

 Detector level samples were generated for a default set of physics
 processes at all centre-of-mass energies considered.  Additional
 samples were generated for systematic studies as described in
 Section~\ref{sec:no-cr_models}.  Hadron level samples were produced for
 all variants of \WW\ events\footnote{In this paper, ``\WW\ events''
 implies doubly-resonant W pair production diagrams, {\em i.e.}
 $t$-channel $\nu_{\mathrm{e}}$ exchange and $s$-channel \Zgamma\
 exchange, referred to as ``\CC'' in \cite{bib:LEP2YR1}.} considered
 (different hadronisation and colour reconnection models) at all
 centre-of-mass energies.  The Monte Carlo event generators used to
 simulate the physics processes are described in the remainder of this
 Section, with emphasis on the CR models themselves.

 The effects of colour reconnection are implemented in several \WW\
 event generators, and three groups of such models are studied, namely
 those of Sj\"ostrand and Khoze (\SK) \cite{bib:SK,bib:SK98}, and
 those implemented in \Ariadne~4.11 \cite{bib:LL,bib:ARIADNE} and in
 \Herwig~6.2 \cite{bib:webbercr,bib:HERWIG}.  Events from all CR
 models used in this paper have been generated in conjunction with the
 electroweak generator \Koralw~1.42 (KW) \cite{bib:KORALW}.  For the
 SK models, samples of events were generated such that they are
 identical to the conventional \Koralw\ \WWqqqq\ events up to the end
 of the parton shower.  This allows the construction of samples with
 an arbitrary fraction of reconnected events (including detector
 simulation) for the \SKI\ model, and also improves the statistical
 precision of studies using the \SK\ models, such as estimation of CR
 bias in measurements of the W boson mass \cite{bib:opalmw_final}.
 For other CR models, the electroweak process was generated using
 \Koralw\ and then a single set of events was hadronised by each of
 \Herwig\ and the \Ariadne\ models.  The same \Koralw\ events were
 also hadronised using the conventional QCD models of
 Section~\ref{sec:no-cr_models}.  All models considered have been
 tuned to describe \Zz\ data, as described in \cite{bib:opalmw_final}.

 \subsection{\SK\ CR Models}
 \label{sec:SK_models}
  The \SK\ models are based upon the Lund string picture of colour
  confinement, in which a string is created that spans the decay
  product partons associated with each W. These strings expand from
  the respective decay vertices and subsequently fragment to hadrons.
  Before this occurs, at most one reconnection is allowed between
  sections of the two strings. The main scenarios considered are
  called type I (\SKI) and type II (\SKII) in analogy to the two types
  of superconducting vortices which could correspond to colour
  strings.  In the \SKI\ scenario, the two colour flux tubes have a
  lateral extent comparable to hadronic dimensions. The probability
  for reconnection to occur is given by $\Prec=(1-\exp(-V\kI))$, where
  $V$ is the space-time integrated product of the maximum colour field
  strengths of the two overlapped W strings and \kI\ is a free
  (dimensionless) strength parameter.  In the \SKII\ scenario, the
  strings have infinitesimally small radii and a unit reconnection
  probability upon their first crossing. A third scenario considered,
  \SKIIpr, is similar to \SKII\ but reconnection is only allowed to
  occur at the first string crossing which would reduce the total
  string length of the system.

  As described in \cite{bib:SK98}, the only tuning necessary for these
  models is to ensure that the \Jetset\ hadronisation model gives a
  good description of \Zz\ data: parameters governing the behaviour of
  the reconnection model are not adjusted to fit data. Therefore, the
  same parameters were used as for the corresponding sample of
  non-reconnected \eeWW\ events. The parton shower cut-off parameter,
  $Q_0$, to which the predictions of the \SKII\ and \SKIIpr\ models in
  particular are sensitive, is set to 1.9~\GeV\ in the OPAL tune
  \cite{bib:opaljttune} of the \Jetset\ hadronisation model.  The
  fractions of \WWqqqq\ events in which reconnection occurs at
  $\roots=199.5$~\GeV, \Prec, are predicted to be 17.2\% for \SKII\
  and 16.1\% for \SKIIpr.  As the fraction of events reconnected
  varies with \kI\ in the \SKI\ model, two illustrative values of \kI\
  are given for comparison in figures and tables: $\kI=0.9$, giving a
  fraction of reconnected events, $\Prec\simeq34.3$\%, comparable to
  that used in \cite{bib:SK}, and an extreme case of $\kI=100$
  ($\Prec\simeq98$\%).  The latter will be referred to hereafter as
  \SKI\ with 100\% CR.

  Samples of these three models including simulation of the detector
  were generated at $\roots = 188.6$~\GeV, 199.5~\GeV\ and 206.0~\GeV.

 \subsection{\Ariadne\ CR Models}
 \label{sec:AR_models}
 The second set of CR models is contained in the \Ariadne\ Monte Carlo
 program. They may be considered as extensions of the earlier partonic
 dipole model\footnote{In \cite{bib:GH}, at most one reconnection was
 allowed per event and possible reconnections between the decay
 products of a single W were not implemented.} \cite{bib:GH}, as both
 models were implemented using the \Ariadne\ Monte Carlo program and
 the same criterion is employed in the reconnection ansatz to
 determine whether reconnection is allowed.  Perturbative QCD favours
 configurations with minimal string length in hadronic \Zz\ decays
 \cite{bib:LEP2YR1}. When the partons of two W bosons are separating
 and strings are being formed between them, it is plausible that
 configurations corresponding to a reduced total string length are
 favoured. In the reconnection model of \Ariadne, the string length is
 defined in terms of the $\Lambda$ measure, which may be viewed as the
 rapidity range along the string: $\Lambda=\sum_i\ln(m_i^2/m_\rho^2)$,
 where $m_i$ is the invariant mass of string segment $i$ and $m_\rho$
 sets a typical hadronic mass scale. Reconnections are only permitted
 if they satisfy the constraints of colour algebra and also lead to a
 reduction in the total $\Lambda$ of the system.  The first model is a
 variant of \Ariadne\ in which rearrangement of the colour flow is
 allowed but is restricted to the decay products of each W separately.
 This is referred to as \ARI.  The second \Ariadne\ model, referred to
 herein as \ARII, is the same as \ARI\ but in addition allows
 reconnections between the two W bosons for gluons having energies, $E_g <
 \Gw$, while the third \Ariadne\ model, \ARIII, does not impose such a
 restriction.  As gluons emitted with $E_g > \Gw\sim 2$~\GeV\ are
 perturbative in nature and have been shown to be radiated
 incoherently by the two initial colour dipoles \cite{bib:SK}, the
 model \ARIII\ is disfavoured on theoretical grounds.  In addition,
 the \Ariadne\ colour reconnection models have been shown to be
 disfavoured by \Zqq\ data \cite{bib:CR_JWG}.  The \ARIII\ model is
 therefore not considered further in this paper.

 The way in which CR is implemented in \ARII\ leads to an artificial
 difference relative to the \ARI\ model which is not directly related
 to reconnections between the two W systems.  In \ARII, the dipole
 cascade (ordering in transverse momentum) is run in two stages from
 the maximum allowed gluon energy down to the cutoff: once down to
 $E_g=\Gw$ allowing only reconnections within a single W system, and
 then a second time allowing $E_g<\Gw$ and cross-talk between the
 two W systems.  In \ARI, the dipole cascade is carried out in a
 single stage without any interruption down to the cutoff.  As noted
 in \cite{bib:LL}, the \ARII\ scheme is not strictly consistent
 with the assumptions of ordering in transverse momentum in the dipole
 cascade model and this leads to the observable differences between \ARII\ and
 \ARI\ referred to above.  To ensure that differences between these
 two models are only due to inter-W reconnections, the dipole cascade
 in \ARI\ is modified to run in two stages with an interruption at
 $E_g=\Gw$ \cite{bib:ar_2phase}.

 As the same tuning of model parameters is used for both \ARI\ and
 \ARII, and no colour flow takes place between the two W bosons in
 events in the \ARI\ model, the \ARI\ model serves as the no-CR model
 when estimating the expected sensitivity of the measurements to
 colour reconnection.  The \ARI\ model is also used as an alternative
 model when estimating systematic effects in the hadronisation of the
 W decay products.

 Samples of the \ARI\ and \ARII\ models including simulation of the
 detector were generated at $\roots = 199.5$~\GeV, with the fraction
 of \WWqqqq\ events in which reconnection occurs being approximately
 49.4\% for \ARII.

 \subsection{\Herwig\ CR Model}
 \label{sec:HW_models}
 The third model is contained in the \Herwig\ program and provides an
 alternative CR model based on cluster hadronisation.  In the cluster
 model, quarks and gluons from the perturbative parton shower
 evolution are combined locally into colour singlet objects called
 clusters which have (relative to strings) low mass and small
 space-time extent, each cluster decaying directly into a small number
 of hadrons.  In the CR version of this model, an alternative pattern
 of cluster formation is implemented after the parton shower and gluon
 splitting phase.  In this, new associations of partons into clusters
 are considered where they would lead to a smaller space-time extent
 of the clusters.  When such viable alternative parton-cluster
 associations exist, they occur with a probability equal to
 $\frac{1}{9}$ ($= 1/N_{\mathrm{colours}}^2$).  A sample of events
 including simulation of the detector was generated at
 $\roots=199.5$~\GeV, with reconnection occurring in approximately 23\%
 of \WWqqqq\ events.

 \subsection{No-CR Monte Carlo Samples}
 \label{sec:no-cr_models}
 The models used are the same as those in \cite{bib:opalmw_final},
 where a more detailed description can be found, and all are generated
 at the detector level.  Samples of \WW\ events without colour
 reconnection effects are simulated at all centre-of-mass energies
 using the \Koralw\ Monte Carlo, with fragmentation carried out using
 the \Jetset~7.408 model.  At three centre-of-mass energies
 (188.6~\GeV, 199.5~\GeV, 206.0~\GeV), additional samples are used in
 which the underlying \WW\ production process is simulated by \Koralw,
 while the fragmentation of a given set of four fermions is performed
 by either \Herwig, \Ariadne\ or an older parameter set of \Jetset\
 derived from tuning the model to \Opal\ inclusive event shape data,
 as used in \cite{bib:opalmw189}.  Hereafter, these samples are
 referred to as \Jetset, \Herwig, \Ariadne\ and ``old \Jetset'',
 respectively.  Samples of \WW\ events including Bose-Einstein
 correlations are simulated using the LUBOEI model \cite{bib:luboei}
 in \Pythia~6.125 \cite{bib:PYTHIA}.
 
 The dominant backgrounds in the \WWqqqq\ channel are \epem\
 $\rightarrow$ \Zqq\ with radiation of energetic gluons, and
 four-fermion final states, primarily \epem\ $\rightarrow$ \ZZqqqq.
 Backgrounds in the \WWqqlv\ channel are significantly lower from all
 sources and receive small additional contributions from \epem\
 $\rightarrow$ \Ztt, simulated using the \KKff\ version 4 Monte Carlo
 program \cite{bib:kkff}.  For completeness, the small background
 represented by hadronic two-photon events is simulated using the
 \PHOJET\ \cite{bib:PHOJET} and \Herwig\ event generators.

 Four samples of two-fermion processes
 $\epem\rightarrow\Zgamma\rightarrow\qq$ are simulated at each
 centre-of-mass energy, to allow systematic uncertainties to be
 estimated: in three samples the hard process is generated using
 \KKff, with fragmentation of the quarks performed by each of \Jetset,
 \Herwig\ and \Ariadne, while in a fourth sample both the hard process
 and fragmentation are generated by \Pythia~6.125.

 Four-fermion processes are modelled using the \Koralw~1.42 Monte
 Carlo, which contains matrix elements calculated by \grcff~2.0
 \cite{bib:GRC4F}.  The complete four-fermion samples are divided into
 two categories: ``WW-like four-fermion events'', corresponding to final
 states which could have been produced by diagrams involving at least
 one W boson, and ``ZZ events'', which are the complementary sample,
 not all of which involve two Z bosons.

  In this and similar analyses, \ZZqqqq\ events are considered as
  background.  In general, the susceptibility to the effects of CR in
  such events is expected to be comparable to that in \WWqqqq.  A more
  complete treatment of such events would require implementation of
  the CR models in a full four-fermion generator, thereby obviating the
  need explicitly to subtract ZZ four quark final states.  Although
  this is not done, it is expected to have only a small effect as the
  background level from \ZZqqqq\ events is low (less than 5\% in the
  particle flow analysis).

  This four-fermion ``background'' is constructed from the
  difference between the predictions of two classes of events
  generated using the \Koralw\ model: one containing the full set of
  interfering four-fermion diagrams (WW-like four-fermion events and ZZ
  events), the other containing only the W pair production diagrams.

 Alternative modelling of the four-fermion process with a more
 complete treatment of so-called $O(\alpha)$ photon radiation has
 been studied using samples generated by the \Kandy\ generator
 \cite{bib:KORALW151} which incorporates \Koralw~1.51
 \cite{bib:KORALW151} and YFSWW3 \cite{bib:YFSWW}.

\section{Data Analysis and Correction Procedure}
\label{sec:analysis}
  
 The measurements of the inclusive charged particle multiplicity and
 of the particle flow are described below.  The former is a
 fully inclusive measurement and data are corrected for the effects of
 finite detector resolution and acceptance, whereas the latter
 compares the predictions of a variety of models with the data at
 detector level.

 \subsection{Inclusive Charged Particle Multiplicity}
 \label{sec:nch-method}
 The analysis of charged particle multiplicity follows the unfolding
 procedure described in \cite{bib:opalcr183}.  The distributions of
 particle multiplicity and of the scaled charged particle momentum,
 $\xp=p/\Ebeam$, where \Ebeam\ is the beam energy, are used to measure
 the mean charged particle multiplicities in \WWqqqq\ events
 (\nchQQQQ) events, and in \WWqqlv\ events (\nchQQLV), and their difference
 ($\Dnch=\nchQQQQ-2\nchQQLV$).

\begin{figure}[htbp]
 \centerline{\epsfig{file=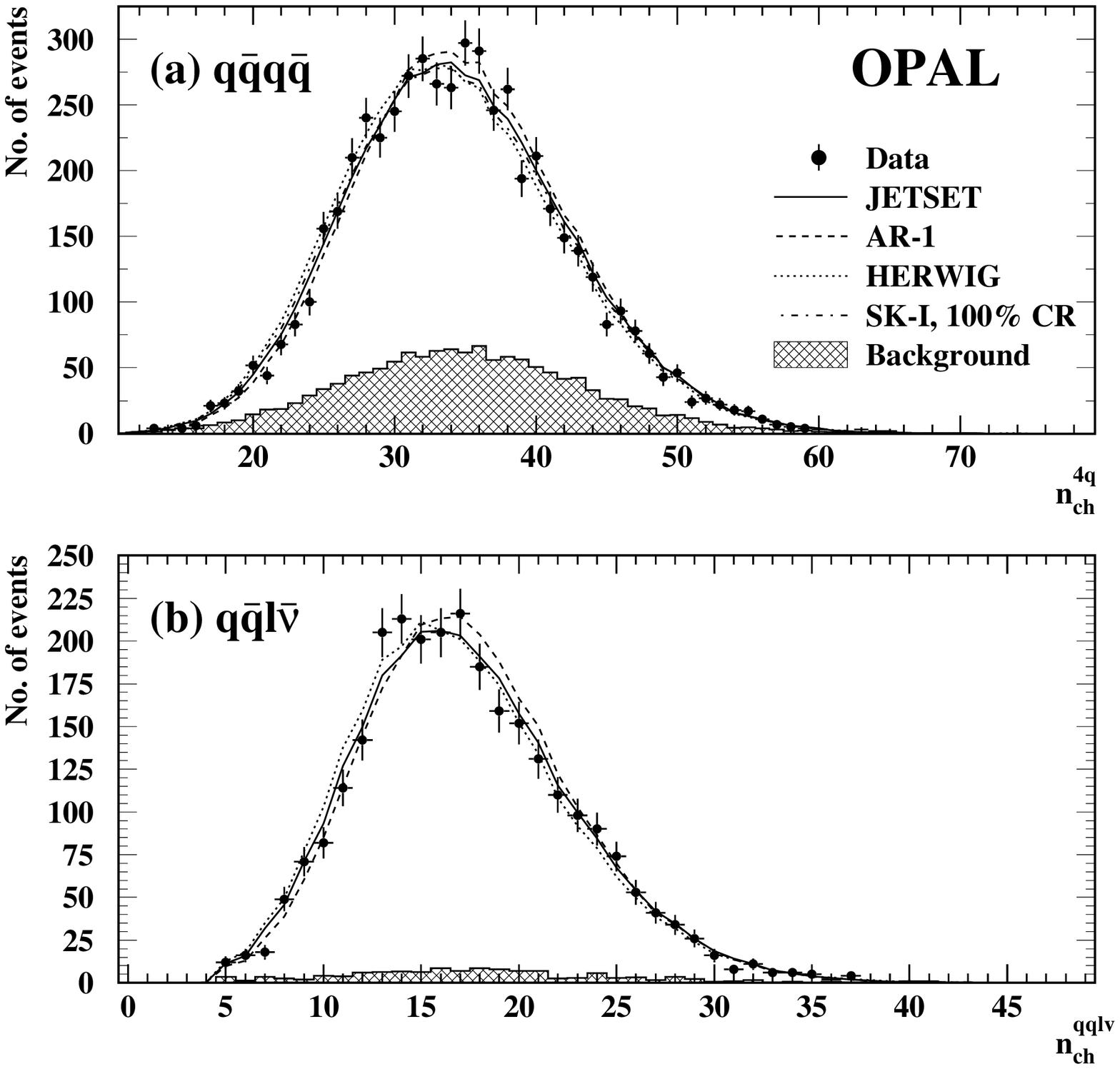,width=\textwidth}}
 \caption{Uncorrected charged particle multiplicity distributions for
 data in the range \roots=189--209~\GeV: (a) \WWqqqq\ events and (b)
 the hadronic part of \WWqqlv\ events.  Points indicate the data with
 statistical errors, lines show the expected sum of signal and
 background contributions for a variety of signal models, and the
 hatched histogram shows the expected background. Predictions of the
 conventional QCD hadronisation models \Jetset\ and \Herwig, the \ARI\
 model and the 100\% CR \SKI\ model, are shown.}
 \label{fig:nch_dists}
\end{figure}

\begin{figure}[htbp]
 \centerline{\epsfig{file=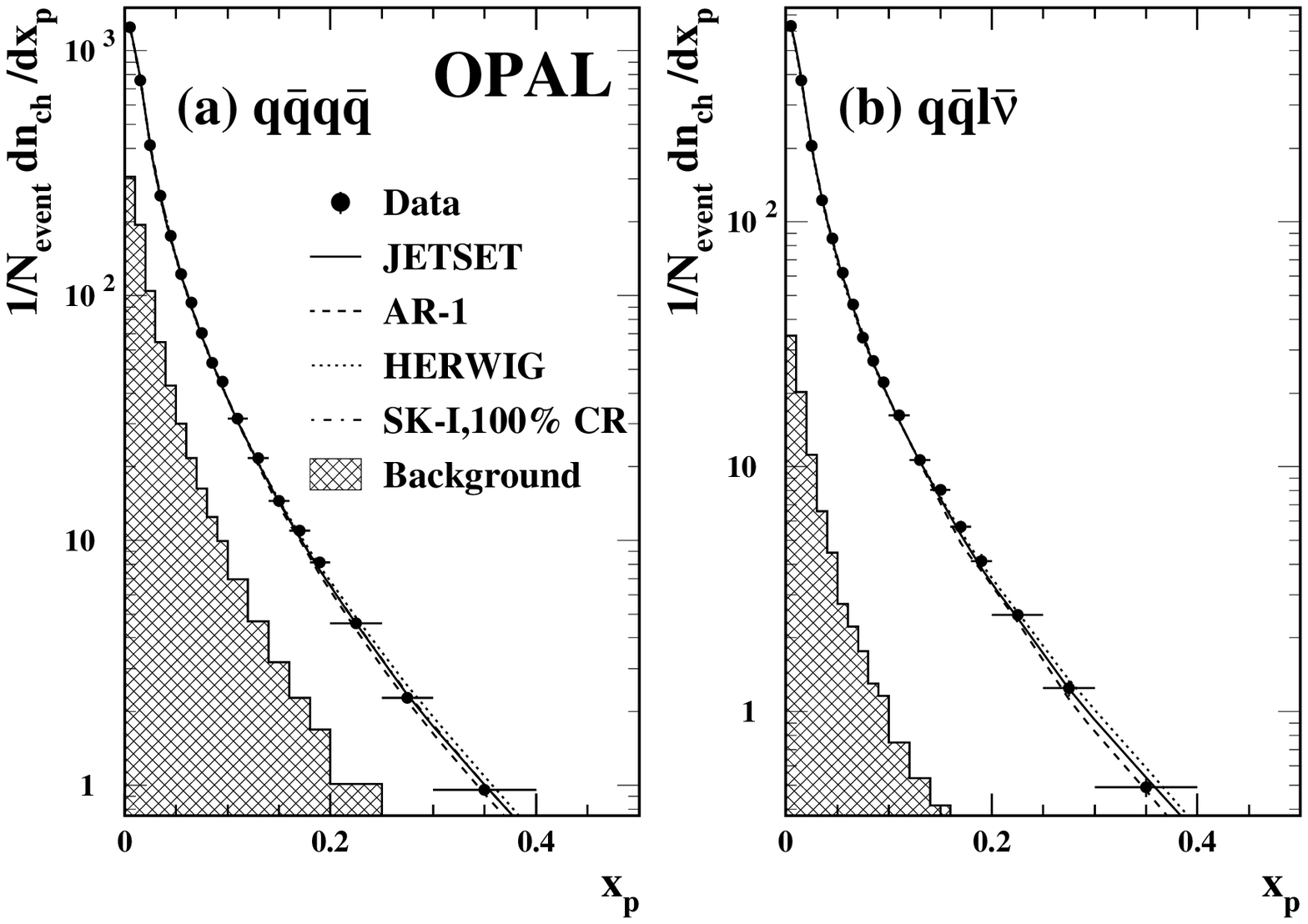,width=\textwidth}}
 \caption{Uncorrected \xp\ distributions for data in the range
 \roots=189--209~\GeV: (a) \WWqqqq\ events and (b) the hadronic part
 of \WWqqlv\ events.  Points indicate the data with statistical
 errors, smooth curves show the expected sum of signal and background
 contributions for a variety of signal models, and the hatched region
 shows the expected background. Predictions of the conventional QCD
 hadronisation models \Jetset\ and \Herwig, the \ARI\ model and the
 100\% CR \SKI\ model, are shown.  Monte Carlo samples are normalised
 to the predicted number of signal plus background events, therefore
 the hatched regions correspond to the mean number of particles in
 candidate \WW\ events which originate from background sources, rather
 than the mean number of particles per background event.}
 \label{fig:xp_dists}
\end{figure}

  Figures~\ref{fig:nch_dists} and \ref{fig:xp_dists} show the
 uncorrected multiplicity and \xp\ distributions for \WW\ candidate
 events before background subtraction.  The background predictions are
 the sum of all other Standard Model processes, as described by the
 models outlined in Section~\ref{sec:no-cr_models}.  The data are
 described reasonably well by all \WW\ models including conventional
 QCD processes alone, and by those including CR\@.  Integration of the
 \xp\ distribution is used for the principal measurement of mean
 charged particle multiplicity as it has slightly lower estimated
 systematic effects than the direct multiplicity measurement, which is
 therefore used as a cross-check.

 The \xp\ distribution is corrected for contamination using a
 bin-by-bin subtraction of the expected background, based on Monte
 Carlo estimates.  Corrections are then applied for finite acceptance
 and the effects of detector resolution, using two samples of \eeWW\
 events generated using the same Monte Carlo event generator at the
 same $\sqrt{s}$, one at hadron level, the other at detector level.
 Distributions normalised to the number of events at the detector and
 the hadron level are compared to derive bin-by-bin correction factors
 which are used to correct the observed \xp\ distribution at each
 centre-of-mass energy.

 This bin-by-bin unfolding procedure is suitable for \xp\ as the
 effects of finite resolution and acceptance do not cause significant
 migration (and therefore correlation) between bins.  Such a method is
 not readily applicable to multiplicity distributions, due to the
 large correlations between bins.  Instead, a matrix correction is
 used to correct for detector resolution effects, followed by a
 bin-by-bin correction which accounts for the residual effects due to
 acceptance cuts and initial state radiation, as in previous \Opal\
 multiplicity studies, e.g. \cite{bib:LEP1.5QCD,bib:opalcr183}.

\begin{figure}[htbp]
 \centerline{\epsfig{file=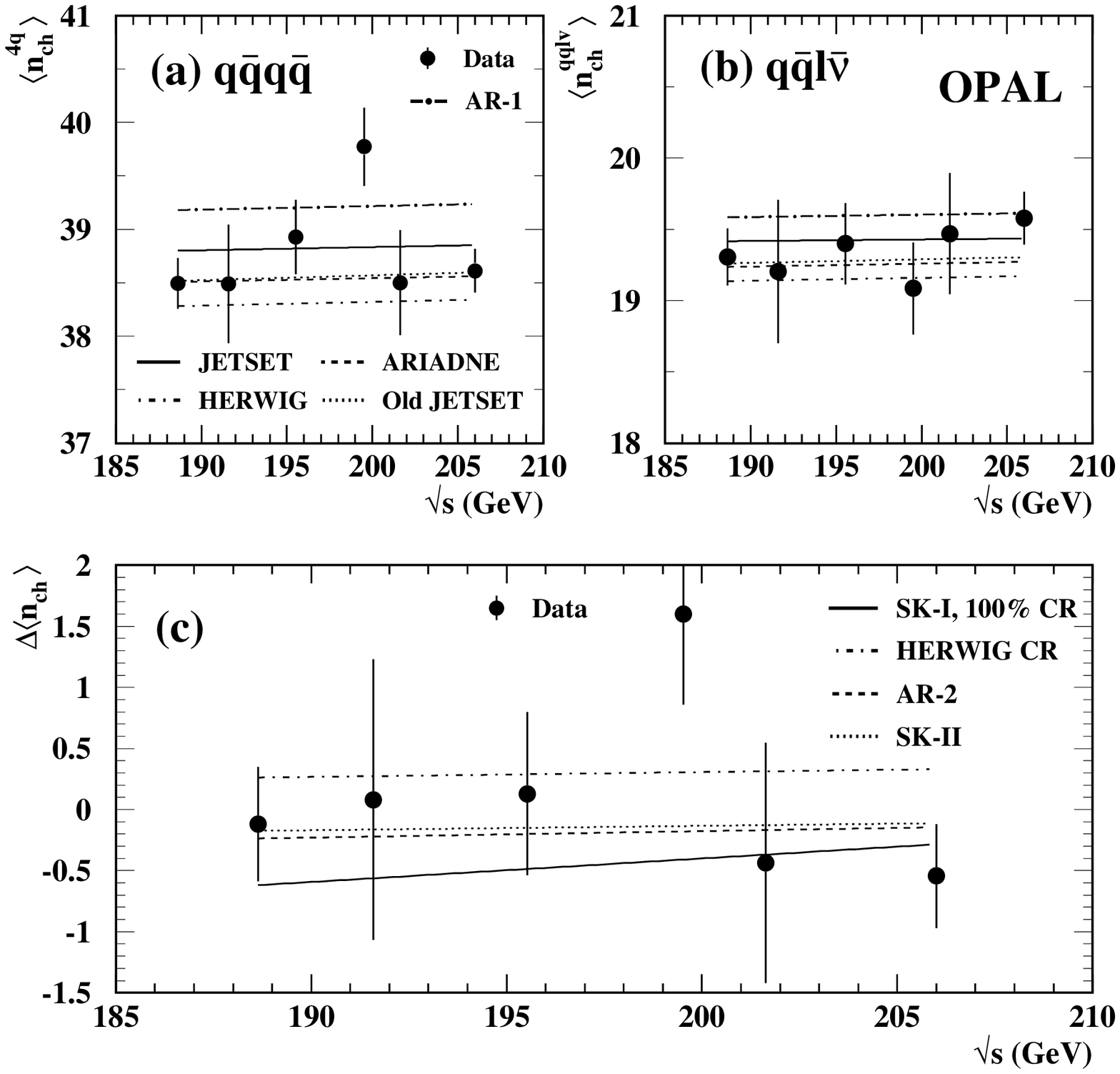,width=\textwidth}}
 \caption{Centre-of-mass energy dependence of the measured (unfolded)
 mean charged particle multiplicity for (a) \WWqqqq\ events, (b)
 \WWqqlv\ events, and (c) the difference, $\Dnch=\nchQQQQ-2\nchQQLV$.
 Points indicate the data with statistical errors and lines the
 predictions of \WW\ models incorporating either conventional QCD
 hadronisation or CR.  The predictions of \Jetset, \Herwig, \Ariadne,
 the old tune of \Jetset\ and \ARI\ are indistinguishable from zero in
 (c), in all cases having values smaller in magnitude than 0.05, and
 so are not shown.}
 \label{fig:nch_roots}
\end{figure}

 Figure~\ref{fig:nch_roots} shows the corrected values of mean charged
 particle multiplicity, \nchQQQQ, \nchQQLV\ and their difference
 \Dnch\ as a function of centre-of-mass energy.  It can be seen that,
 although the no-CR models vary in their predictions for both
 \nchQQQQ\ and \nchQQLV, they are in complete agreement that the value
 of \Dnch\ is negligibly small, in contrast to the CR models shown in
 Figure~\ref{fig:nch_roots}(c).  However, the predictions of both
 conventional QCD models and models of CR are found to be compatible
 with the data within uncertainties.  As the multiplicity data are not
 observed to vary significantly with \roots, measurements from all
 centre-of-mass energies are combined assuming they are independent of
 \roots.  The combined results are presented in
 Table~\ref{tab:systematics_nch}, systematic uncertainties discussed
 in Section~\ref{sec:systematics}, and quantitative comparisons with
 models presented in Table~\ref{tab:results}.

 \subsection{Particle Flow}

  The analysis of event properties presented here is a generalisation
  of the string effect analysis in three-jet \eeqqg\ events to the
  four-jet topology of \WWqqqq. The situation is necessarily more
  complicated in the \WWqqqq\ channel because, in contrast to the
  three-jet case, events are not constrained by momentum conservation
  to be planar.  The analysis is therefore carried out in four
  distinct planes, each of which is defined by a pair of jet axes.
  Charged particles and clusters of electromagnetic energy, selected
  as in \cite{bib:opalcr183}, are combined into four jets using the
  $k_\perp$ \cite{bib:durham} jet-finding algorithm, and the total
  momentum and energy of each of the jets are corrected empirically
  for double counting using the same energy flow algorithm employed in
  \cite{bib:opalmw_final}. The jet momenta are further modified by a
  kinematic fit, imposing the four constraints of energy and momentum
  conservation, to obtain an improved estimate of the trajectories of
  the underlying four fermions from the \WW\ decays.  To ensure that a
  relatively simple colour topology is being studied, events having a
  five-jet like topology\footnote{Following \cite{bib:opalmw_final},
  five-jet events are classified as those in which the $k_\perp$ jet
  resolution parameter for the four-jet to five-jet transition,
  $y_{45}$, satisfies $\log(y_{45})>-5.6$.}  are rejected.  This is
  also expected to lead to a better description of the dominant \eeqq\
  background events by the parton shower models.

  The analysis proceeds in three stages, namely: association of pairs
  of jets with W bosons and definition of four planes, projection of
  charged particles onto these planes, and comparison of the distributions of
  particles in these planes. Each of these aspects of the analysis is
  described below.

  The association between pairs of jets and W bosons is performed using a
  minor variant of the algorithm that was introduced in
  \cite{bib:opalmw189}. In the current scheme, the mass obtained from
  a five-constraint\footnote{The additional constraint imposed is
  equality of the masses of the two W boson candidates.} kinematic fit
  \cite{bib:opalmw_final} is combined with the variables of
  \cite{bib:opalmw189} into a single likelihood discriminant,
  selecting the correct pairing of jets to W bosons with a purity of
  $\sim90$\%.  The total number of events used in the analysis after
  all selections is 2199, with an overall efficiency for selecting
  \WWqqqq\ of $\sim40$\%.  Small variations in performance with
  centre-of-mass energy are detailed in Table~\ref{tab:data_summary}.

\begin{figure}[htbp]
 \centerline{\epsfig{file=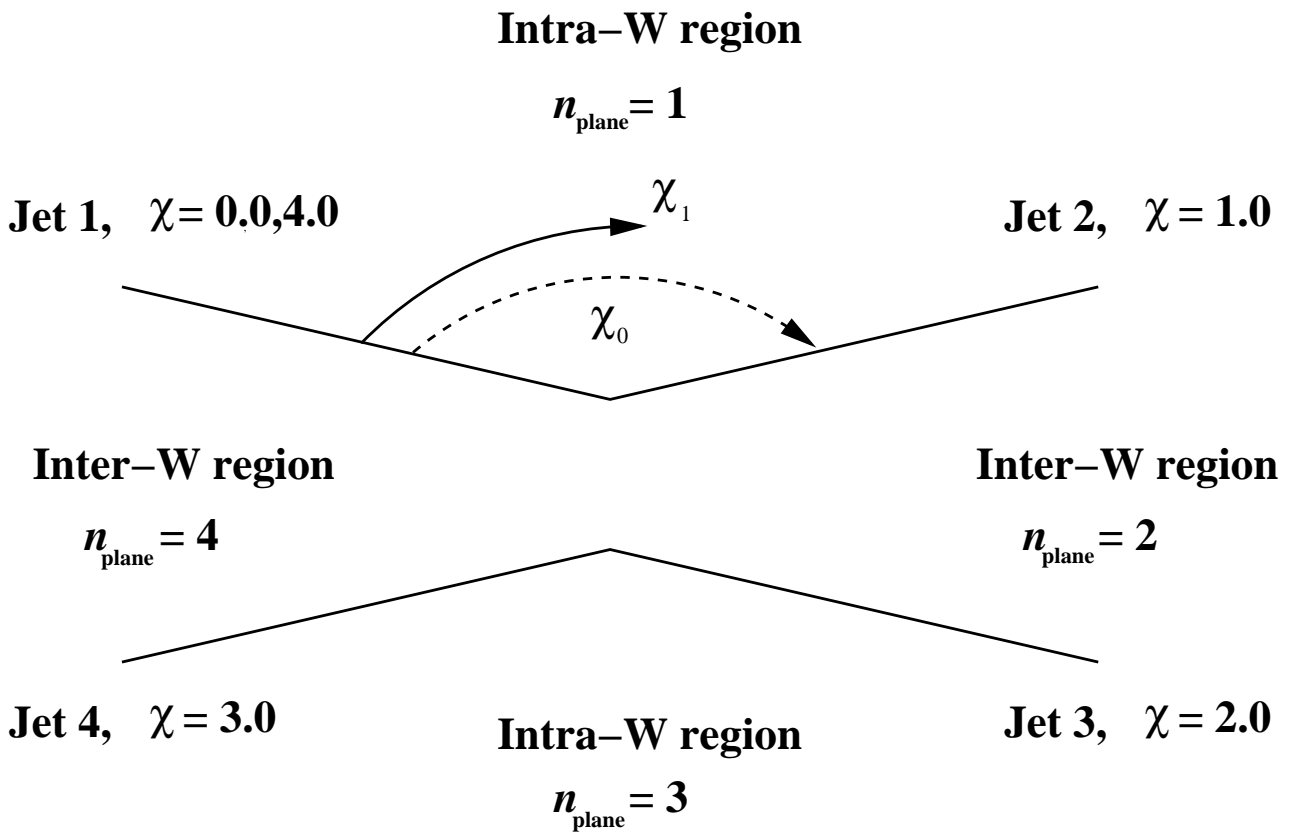,width=\textwidth}}
 \caption{Illustration of inter-W and intra-W regions, and numbering
 convention adopted, as described in Section~\ref{sec:analysis}.}
 \label{fig:topology_sketch}
\end{figure}

  The pairing of jets originating from the same parent W defines two
  intra-W planes, as illustrated in Figure~\ref{fig:topology_sketch}.
  With four jets, there exist two ways in which planes may be defined
  between jets which originate from different W boson (inter-W regions).
  The configuration which results in the smaller sum of inter-W angles
  is chosen.  The motivation for this is the suggestion
  \cite{bib:SK,bib:LL,bib:GH} that colour reconnection is more
  probable for topologies in which jets from different W bosons are close
  together in angle.  In such a configuration, a rearrangement of the
  colour flow in the event would be energetically favoured due to a
  reduction in the overall ``length'' of the colour flux tubes.

  Reconstructed charged particles in the event are projected onto the
 intra-W and inter-W planes as follows, and illustrated in
 Figure~\ref{fig:topology_sketch}.  The first plane examined is that defined
 by the most energetic jet in the event (`jet 1') and the jet belonging to
 the same W (`jet 2'), as given by the jet pairing algorithm.  The next
 plane considered is that between jet 2 and a jet (`jet 3') from the other
 W, such that the sum of inter-W angles is minimal.  The third plane is the
 other intra-W region between jet 3 and the remaining jet in the event, `jet
 4'.  The final plane is that between jet 4 and jet 1.

  All charged particles in an event are projected onto each of the
  four planes in turn.  In each plane, an azimuthal angle
  $0<\chi_1<2\pi$ is defined, having positive sense between pairs of
  jets as described above and indicated in
  Figure~\ref{fig:topology_sketch}.  To account for the variation in
  the angle between pairs of jets, the distribution of particles is
  evaluated as a function of $\chi_1$ for each plane after rescaling,
  event-by-event, to the angle between the jets which define the
  plane, $\chi_0$.  This gives a normalised angle,
  $\chi_R=\chi_1/\chi_0$, where $\chi_R\equiv0$ corresponds to the jet
  axis of the lower number jet which defines the plane.  Particles
  outside the inter-jet region, \ie, having $\chi_R>1$, are not
  considered further.  In the case where a particle is projected into
  the inter-jet region of more than one plane, it is exclusively
  assigned to the plane relative to which it has the smallest
  transverse momentum.

\begin{figure}[htbp]
 \centerline{\epsfig{file=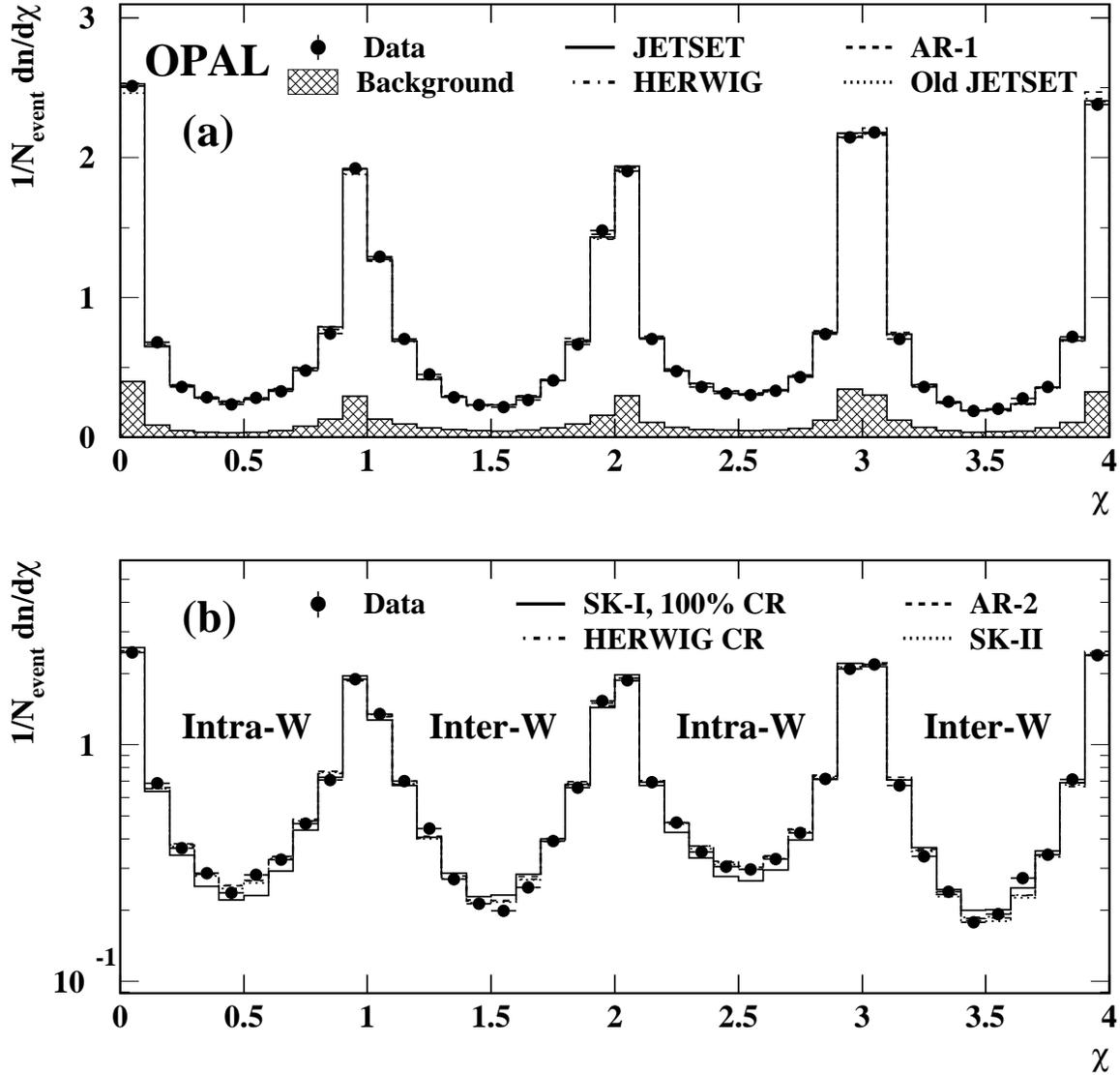,width=\textwidth}}
 \caption{The particle flow distribution in the four inter-jet planes,
 normalised event-by-event to the inter-jet angles, as described in
 the text and in Figure~\ref{fig:topology_sketch}.  Points represent
 the data with statistical errors. (a) compares the data with the
 predictions of conventional QCD hadronisation Monte Carlo models and
 \ARI, (b) compares the data, after background subtraction, with
 several CR models. Monte Carlo samples are normalised to the
 predicted number of signal plus background events, therefore the
 hatched region in (a) corresponds to the mean number of particles in
 candidate \WWqqqq\ events which originate from background sources,
 rather than the mean number of particles per background event.}
\label{fig:flow_reduced}
\end{figure}

  The four normalised inter-jet regions are combined in a single
  distribution in the range $0<\chi<4$, as shown in
  Figure~\ref{fig:flow_reduced}(a),
  where the structure of the four jets is apparent.  The
  variable $\chi$ is defined as $\chi=\chi_R+(n_{\mathrm{plane}}-1)$,
  where $n_{\mathrm{plane}}$ is an integer between 1 and 4,
  corresponding to the four planes in the order given, and as shown in
  Figure~\ref{fig:topology_sketch}.

   As seen in Figure~\ref{fig:flow_reduced}(a), the data are
   consistent with the predictions of \WW\ production using the
   conventional hadronisation models plus the sum of all background
   processes.  Figure~\ref{fig:flow_reduced}(b) compares the sum of
   background-subtracted data with predictions from various CR models.
   The data are found to be adequately described by models, with the
   exception of the extreme scenario of the \SKI\ model, which
   predicts lower particle densities in the intra-W regions, and
   higher particle densities in the inter-W regions.

  In conventional QCD models without interaction of the colour fields
  between the \Wp\ and \Wm, the particle density (or particle flow) is
  expected to be higher in the intra-W regions, $0<\chi<1$ and
  $2<\chi<3$ than in the inter-W regions.  After a rearrangement, in
  addition to a change in absolute number of charged particles in the
  event, there may be a migration of particle flow away from these
  regions in favour of the inter-W regions, $1<\chi<2$ and $3<\chi<4$.
  Consequently, one way of studying the effects of colour
  rearrangement is to compare the particle flow of the two intra-W
  regions to that of the two inter-W regions.  As the properties of
  the two inter-W regions should both be affected by colour
  rearrangement in the same way, these are added together, as are the
  two intra-W regions.  The ratio
  of intra-W to inter-W particle flow distributions is then
  formed,
\begin{displaymath}
   \Rflow =
\frac{\frac{\mathrm{d}\nch}{\mathrm{d}\chi_R} (\mathrm{intra\!-\!W})}
     {\frac{\mathrm{d}\nch}{\mathrm{d}\chi_R} (\mathrm{inter\!-\!W})},
\end{displaymath}
 where \nch\ is the number of charged particle tracks projected into a
 given inter-jet region.

\begin{figure}[htbp]
 \centerline{\epsfig{file=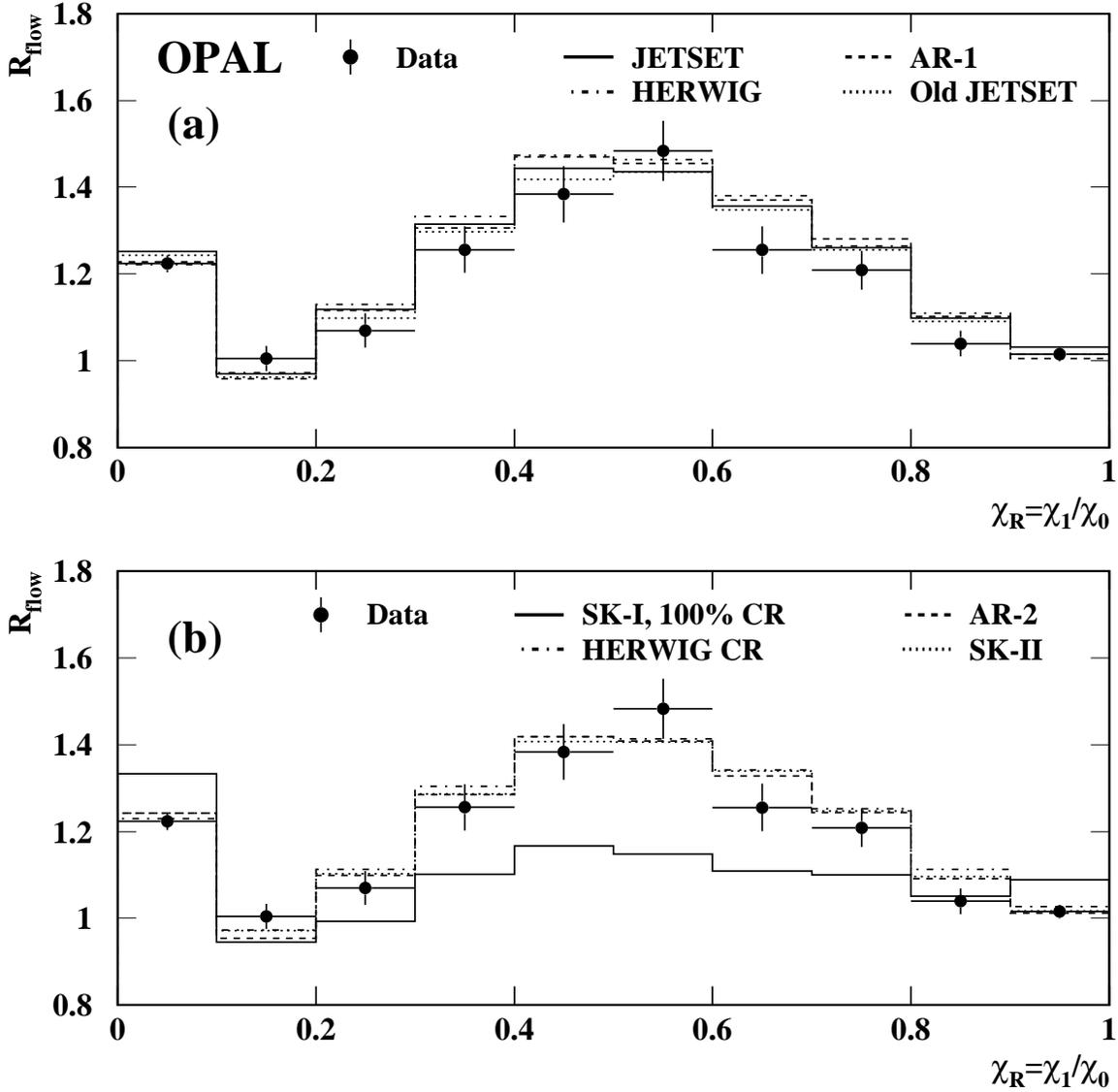,width=\textwidth}}
 \caption{Distribution of \Rflow, the ratio of the particle flow in
 the intra-W regions to that in the inter-W regions.  Points represent
 the data after background subtraction, with statistical errors. (a)
 compares the data with the predictions of conventional QCD
 hadronisation Monte Carlo models and \ARI, (b) compares the data with CR
 models. Note that there are correlations between bins in these
 distributions.}
 \label{fig:flow_ratio}
\end{figure}

  Figure~\ref{fig:flow_ratio}(a) compares the measured values of this
 ratio using all data (after background subtraction) with the
 predictions of conventional hadronisation models, while
 Figure~\ref{fig:flow_ratio}(b) shows data compared with the
 predictions of various CR models.

  Differences between the data and models, and consistency between the
 predictions of different models, are more apparent in
 Figure~\ref{fig:flow_ratio} than in
 Figure~\ref{fig:flow_reduced}. The data are found to lie slightly
 below the model predictions in the region away from the jet cores for
 conventional QCD Monte Carlo models and most CR models.  While the
 \SKI\ model with $\kI=100$ shows significant separation from data and
 the other models, it is apparent that the predicted effects of CR in
 the other \SK\ models are limited.

 \subsubsection{Quantitative Measures of CR}
 \label{sec:quantitative} To quantify the consistency between data and
 predictions of models, the ratio of the integral of the particle flow
 in the intra-W regions to the integral of particle flow in the
 inter-W regions,
\begin{equation}
   \Rn =
\frac{ \int^{0.8}_{0.2} \frac{\mathrm{d}\nch}{\mathrm{d}\chi_R}
                            (\mathrm{intra\!-\!W}) \mathrm{d}\chi_R}
     { \int^{0.8}_{0.2} \frac{\mathrm{d}\nch}{\mathrm{d}\chi_R}
                            (\mathrm{inter\!-\!W}) \mathrm{d}\chi_R} \,\,
 \label{eq:Rn}
\end{equation}
 is formed.  This is a traditional observable used in string effect
  studies, \eg\ \cite{bib:string_effect}, and is sensitive to
  differences in the number of particles in the inter-jet regions but
  relatively insensitive to their angular distribution and the choice
  of binning.

  The limits of integration are chosen to optimise the predicted
 sensitivity in the \SKI\ model at \roots=189~\GeV.  This choice also
 allows the uncertainty on the ratio to be calculated from error
 propagation.  In the case where the limits are extended too close to
 the cores of the jets, a significant correlation is introduced
 between neighbouring inter-jet regions and the error calculation is
 no longer valid.  It is to be noted that to calculate the uncertainty
 correctly, the numerator and denominator of Equation~\ref{eq:Rn} must
 be evaluated event-by-event, rather than by integration of
 distributions such as Figure~\ref{fig:flow_reduced}.  The validity of
 the statistical errors has been tested using data-sized samples from
 a variety of Monte Carlo models, each with more than 90 times the
 statistics of the entire 189--209~\GeV\ data sample.

  To estimate the sensitivity to colour reconnection effects, the \Rn\
 predicted by each colour reconnection model is compared with that
 obtained from the corresponding ``no reconnection'' scenario of the
 same model.  As a guide to the performance of the analysis, the
 predicted statistical sensitivity of the analysis is summarised in
 Table~\ref{tab:sensitivity} using model predictions at
 $\roots=199.5$~\GeV.  The fraction of reconnected events in each
 model is also shown.  The sensitivity is defined as the difference
 between a given reconnection model and its corresponding ``no
 reconnection'' sample ($\Delta\Rn$), divided by the expected
 statistical uncertainty obtained using all data presented in this
 paper (\sigRnstat).  It can be seen that there is a significant
 sensitivity to the extreme scenario of the \SKI\ model in
 which almost all events are reconnected but limited sensitivity to
 all other CR models considered.

\begin{figure}[htbp]
 \centerline{\epsfig{file=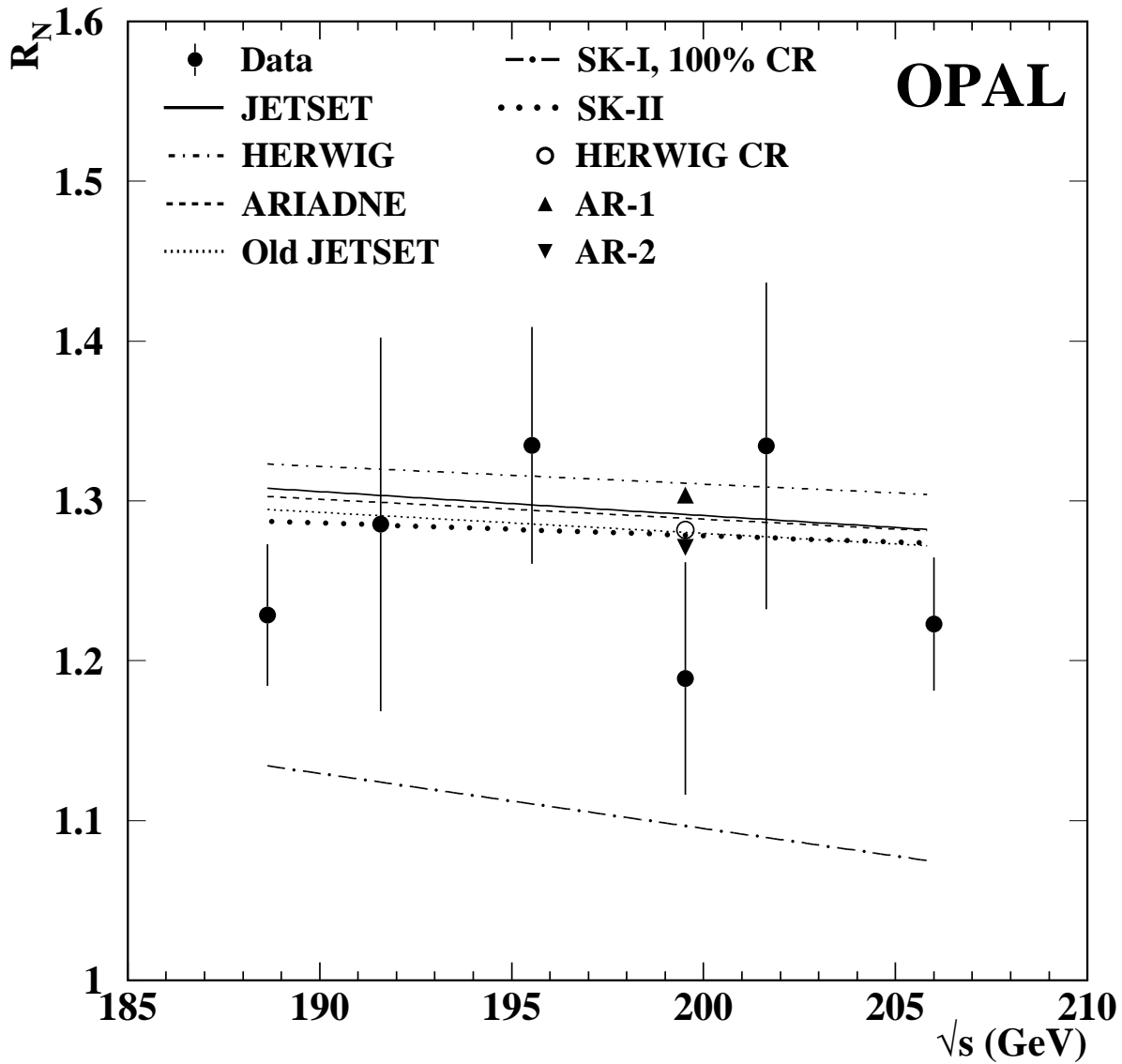,width=\textwidth}}
 \caption{Centre-of-mass energy dependence of \Rn\ for \WWqqqq\
 events.  Points indicate the data with statistical errors and lines the
 predictions of \WW\ models incorporating either conventional QCD
 hadronisation or CR.  The predictions of the \Herwig\ CR model, \ARI\
 and \ARII, which are only available at \roots=199.5~\GeV, are
 indicated by symbols.}
 \label{fig:Rn_roots}
\end{figure}

  To combine the observed \Rn\ from different centre-of-mass energies,
  an assumption has to be made about possible energy dependence of the
  measurements.  Figure~\ref{fig:Rn_roots} shows the measured values
  of \Rn\ together with the predictions of the \Jetset\ and other Monte
  Carlo samples.  Although some models exhibit an energy dependent
  \Rn, the variations are small compared to the statistical
  uncertainties in data.  Measurements are therefore combined assuming
  they are independent of \roots, and the impact of this is considered
  as a systematic effect.  Quantitative comparisons of the combined
  \Rn\ in data are made using predictions of all models studied at
  $\roots=199.5$~\GeV, at which energy a complete set of Monte Carlo
  samples is available.  On the scale of variations predicted in \Rn,
  this \roots\ is close to the luminosity-weighted centre-of-mass
  energy of 197.8~\GeV.

\section{Systematic Uncertainties}
 \label{sec:systematics}

   Systematic uncertainties are studied using measurements averaged
  over \roots, and are shown in Tables~\ref{tab:systematics_nch} and
  \ref{tab:systematics} for the inclusive charged particle
  multiplicity and particle flow analyses, respectively.  Sources of
  systematic error considered include hadronisation effects in the \WW\
  models, detector effects related to tracking of charged particles
  and background subtraction.  The analysis of mean particle
  multiplicities involves the unfolding of observed data to the hadron
  level and a possible additional uncertainty related to this
  procedure is studied.  The particle flow analysis is performed using
  ratios of sums and no unfolding of the data is performed, and so
  many systematic effects are expected to cancel or be negligibly
  small.

 \subsection{\boldmath\WW\ Hadronisation}
 For the multiplicity analysis, samples of simulated \WW\ events
 incorporating \Jetset\ hadronisation are treated as
 background-subtracted data and unfolded with each of the alternative
 \WW\ hadronisation models and colour reconnection models.  The CR
 models are included in this procedure to allow for the possibility
 that such effects are present in the underlying physics. Where
 samples of a given model exist at more than one \roots\ the results
 of unfolding are averaged.  The uncertainty is assigned as the
 largest difference between the values obtained when unfolding
 \Jetset\ events with the default model (\Jetset) and with any of the
 alternative models, and is dominated by the \Herwig\ models.
 
 For the particle flow analysis, this error is assigned by using
 samples of \WW\ events generated using \Koralw, and hadronised with
 each of the models \Jetset, \Herwig, \Ariadne, old \Jetset\ and \ARI.
 Where samples of a given model exist at more than one \roots\ the
 results are averaged.  Each model is treated as background-subtracted
 data, and the uncertainty is assigned as half of the maximum
 difference between the \Rn\ predicted by any pair of models.  This
 differs from the definition used for the multiplicity analysis: as no
 unfolding is performed, there is no default model against which to
 study systematic effects.  It can be seen from
 Figure~\ref{fig:Rn_roots} that this uncertainty is determined by
 differences between the \Herwig\ and old \Jetset\ hadronisation
 models.

 \subsection{BEC}
  While the presence of Bose-Einstein correlations among particles
  originating from the same hadronically decaying W boson (intra-W
  BEC) has been unambiguously established \cite{bib:intra-W_BEC},
  there is no significant evidence for BEC between particles
  originating from different W bosons (inter-W BEC)
  \cite{bib:inter-W_BEC,bib:inter-W_BEC_opal}.
  However, these are not excluded and, following
  \cite{bib:inter-W_BEC_opal} and \cite{bib:opalmw_final}, a
  systematic error corresponding to 77\% of the effect of (inter-W BEC) $-$
  (intra-W BEC) on the measurement is assigned.  In the multiplicity
  analysis, the uncertainty was assigned as 77\% of the difference in
  the hadron level multiplicity obtained when the inter-W BEC and
  intra-W BEC samples were each treated as background-subtracted data.
  In the particle flow analysis the uncertainty was assigned as 77\%
  of the difference between the \Rn\ values predicted by the inter-W
  BEC model and the intra-W BEC model.

 \subsection{Track Definition}

  Uncertainties arising from the selection of charged tracks are
  estimated by examining the stability of the difference between data
  and Monte Carlo predictions for multiplicity or \Rn.  Both analyses
  are repeated three times, with track selection criteria varied
  within reasonable limits \cite{bib:opalcr183}.  The maximum allowed
  values of the distances of closest approach to the interaction
  region in $r$-$\phi$ and $z$ are varied from 2~cm to 5~cm and from
  25~cm to 50~cm, respectively, and the minimum number of hits on
  tracks is varied from 20 to 40.  The uncertainty on the charged
  track definition is the sum in quadrature of these three effects.
  This source represents a significant systematic effect and is
  dominated by the variation of the minimum number of hits required to
  form a track.

 \subsection{Background}
  Alternative models and cross-sections were used to estimate
  uncertainties associated with the background subtraction.  The
  uncertainty is formed using the difference between the measured
  multiplicities (or \Rn\ values) obtained using the alternative
  background models and that obtained using the default
  background model and assumed cross-section at each centre-of-mass
  energy.

  \subsubsection{\boldmath{\eeqq}\ Modelling}
  Uncertainties in generation of the hard process and hadronisation
  may affect the shape of the background and are estimated by comparing
  models.  A sample of \eeqq\ events, generated using \KKff\ and
  hadronised with each of \Jetset, \Herwig\ and \Ariadne, is available
  at all centre-of-mass energies studied, as is a sample generated
  entirely using \Pythia.  This uncertainty is assigned as the largest
  difference between the result obtained using any model and the
  result obtained when the default \Jetset\ model is used.

  \subsubsection{\boldmath{\eeqq}\ Rate}
  This uncertainty arises due to imperfect knowledge of the accepted
  background cross-section.  It is evaluated using the deviations in
  the measurements caused when the \eeqq\ background rate is varied by
  $\pm 5$\% and $\pm 20$\% from its default value for \WWqqqq\ and
  \WWqqlv\ events, respectively, where the allowed ranges are taken
  from \cite{bib:opalxs189}.

 \subsubsection{Four-fermion Background Modelling}

  This systematic uncertainty is estimated by using the \Kandy\
 generator as an alternative to the default (\Koralw) to simulate the
 WW-like four-fermion events, with the change in the measured
 multiplicity or \Rn\ value assigned as the uncertainty.  Note that
 owing to the $O(\alpha)$ corrections in \Kandy, the alternative
 WW-like four-fermion cross-section is 2.5\% lower than that of
 \Koralw, and so the same 2.5\% reduction is also applied to the
 cross-section of the \Koralw\ \WW\ events when carrying out this test.

 \subsubsection{\boldmath{\ZZ}\ Rate}
 An uncertainty is assigned to the assumed cross-section for ZZ
 events.  It is estimated by varying the ZZ component of the four-fermion
 background by $\pm 11$\% for \WWqqqq\ events \cite{bib:opalzz}, and
 $\pm 20$\% for \WWqqlv\ events \cite{bib:opalxs189}.

 \subsubsection{Residual Backgrounds}
   Two further small sources of background are considered.  The first small
   source of background is only relevant for \WWqqlv\ events and so is
   considered for the multiplicity analysis alone.  It is assigned as the
   effect observed on the measurements when all predicted \eell\
   backgrounds are neglected.

  The second source is due to two-photon background, and is estimated as
  the difference found in the final result when the small, predicted
  background from this source is included (default) or neglected.

 \subsection{Unfolding Method}
  For the multiplicity analysis, the results of the \xp\ and direct
  multiplicity analyses are compared and the difference in their
  central values is assigned as a source of possible uncertainty.

 \subsection{Centre-of-Mass Energy Dependence}
  The multiplicity and particle flow measurements are assumed to be
  independent of \roots, as discussed in Section~\ref{sec:analysis}.
  An alternative choice considered for the particle flow analysis was
  to correct the measurements of \Rn\ according to the weak energy
  dependence predicted by the \Koralw\ model with \Jetset\
  hadronisation. The difference between these two assumptions is found
  to be small, at a level of 2\% of the statistical uncertainty on the
  combined result, and is therefore neglected.

 \subsection{Cross-Check Using \boldmath{\WWqqlv}\ Data}
 \label{sec:qqlv_cross_check}
\begin{figure}[htbp]
 \centerline{\epsfig{file=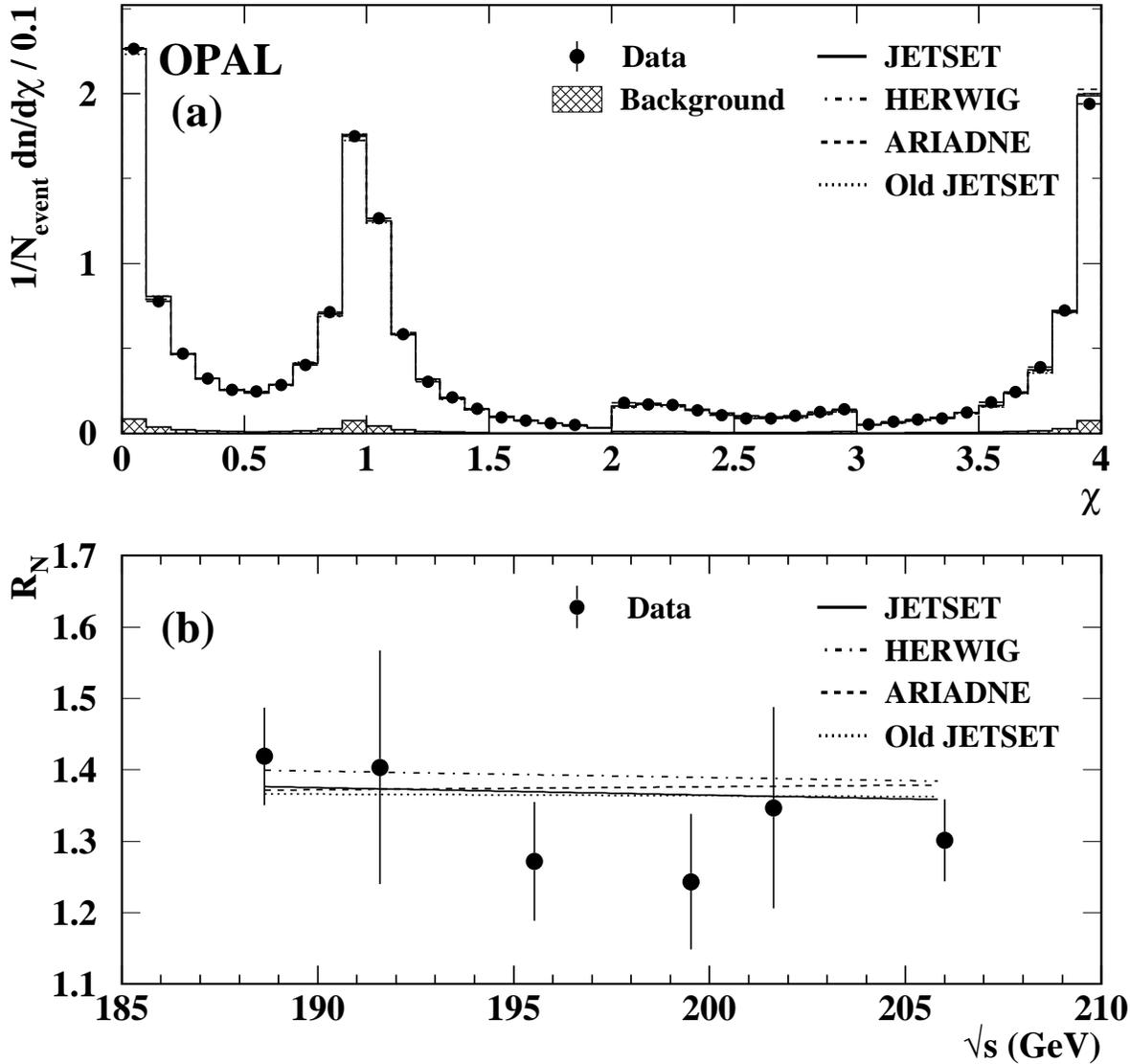,width=\textwidth}}
 \caption{(a) The particle flow distribution (as in
 Figure~\ref{fig:flow_reduced}) for \WWqqlv\ events and conventional
 QCD hadronisation Monte Carlo models, as described in
 Section~\ref{sec:qqlv_cross_check}.  Points represent the data with
 statistical errors and the hatched region the sum of all background
 contributions. (b) The energy evolution of \Rn\ as measured in
 \WWqqlv\ events, in comparison with the predictions of conventional
 QCD models.}\label{fig:Rn_qqlv}
\end{figure}

  By way of a cross-check that the data are adequately described by
  the conventional hadronisation models, the particle flow analysis is
  repeated using \WWqqlv\ events, in which there can be no (inter-W)
  colour reconnection.  The event selection is restricted to events in
  which the charged lepton is classified as either an electron or a
  muon.  The four planes used in these events are defined by the jet
  or fermion directions derived from a kinematic fit in which the
  constraints of energy and momentum conservation are imposed (4-C
  fit, as in \cite{bib:opalmw_final}).  Figure~\ref{fig:Rn_qqlv}(a)
  shows the distribution of particle flow for all \WWqqlv\ data, which
  are found to be described well by the predictions of the \Jetset,
  \Herwig, \Ariadne\ and old \Jetset\ models.

  By construction, the two jet axes corresponding to the hadronically
  decaying W boson are centred at $\chi=0$ and $\chi=1$, while the
  direction of the charged lepton and that inferred for the unobserved
  neutrino are at $\chi=2$ and $\chi=3$, respectively.  Note that
  charged particles associated with the leptonically decaying W bosons
  are not included in these figures.  The non-zero multiplicity in the
  region between the two leptonic ``jets'' is due to the particles
  projected into this plane from the hadronically decaying W.
  Similarly, the abrupt change in the distribution in the region of
  the leptonic W is because none of the charged particles projected
  onto this plane plays a direct role in defining it.

  Figure~\ref{fig:Rn_qqlv}(b) shows the energy evolution of \Rn\ for
  \WWqqlv\ events, together with the predictions of the same set of
  four hadronisation models.  Similarly, the models provide a
  reasonable description of the data, giving confidence to the
  analysis procedure.  No additional systematic uncertainty is
  assigned as a result of this study.

\section{Results and Discussion}
\label{sec:results}

The measurements of the mean charged particle multiplicities corrected
to the hadron level and averaged over the range $\roots\simeq$
189--209~\GeV, are:
\begin{eqnarray*}
           \nchQQQQ & = & 38.74 \pm 0.12 \pm 0.26 \;\; ,\\
           \nchQQLV & = & 19.39 \pm 0.11 \pm 0.09 \;\; ,\\
           \Dnch    & = & -0.04 \pm 0.25 \pm 0.17 \;\; ,
\end{eqnarray*}
 where in each case the first uncertainty is statistical and the
 second systematic.  The difference in mean charged particle
 multiplicities in hadronic W decays in \qq\qq\ and \qq\lnu\ events,
 \Dnch, is found to be consistent with zero within
 uncertainties.  All models considered are found to lie within 1.2
 standard deviations of the measurement, as shown in
 Table~\ref{tab:results}.  As no evidence is found in this measurement
 for colour reconnection between the two hadronically decaying W bosons, the
 average of data from \WWqqqq\ and \WWqqlv\ events, weighted by
 statistical uncertainties and taking into account correlations in the
 systematic uncertainties, is used to yield a measurement of the
 charged particle multiplicity from a single hadronically decaying W,
 \begin{displaymath}
   \nchW    = 19.38 \pm 0.05 \mathrm{(stat.)} \pm 0.08 \mathrm{(syst.)}\;\; .
 \end{displaymath}
As this average is made under the assumption that there is no colour
reconnection between W bosons, the CR contribution to the \WW\ hadronisation
uncertainty has been removed.

 The analysis of particle flow allows a simple comparison with models
 of colour reconnection, using the data of
 Table~\ref{tab:systematics}.  The measurement obtained using
 approximately\ 625~pb$^{-1}$ of data in the range 189--209~\GeV\
 yields:
\begin{equation}
   \Rn = 1.243 \pm0.025 \mathrm{(stat.)} \pm0.023 \mathrm{(syst.)}\;.
\label{eq:rn_avge}
\end{equation}
 This result may be compared with the predictions of the models at
  \roots=199.5~\GeV\ given in Table~\ref{tab:results}.  It can be seen
  that \Rn\ measured in data is lower than all models except the \SKI\
  ($k_I=100$) sample. The (signed) significance of these differences
  is also presented, varying from approximately 4.4 standard
  deviations of the total error (\sigRntot) for an extreme scenario of
  the \SKI\ model, to $-2.0\sigRntot$ for \Herwig, with most other
  models populating a region around $-1\sigRntot$.

\begin{figure}[htbp]
 \centerline{\epsfig{file=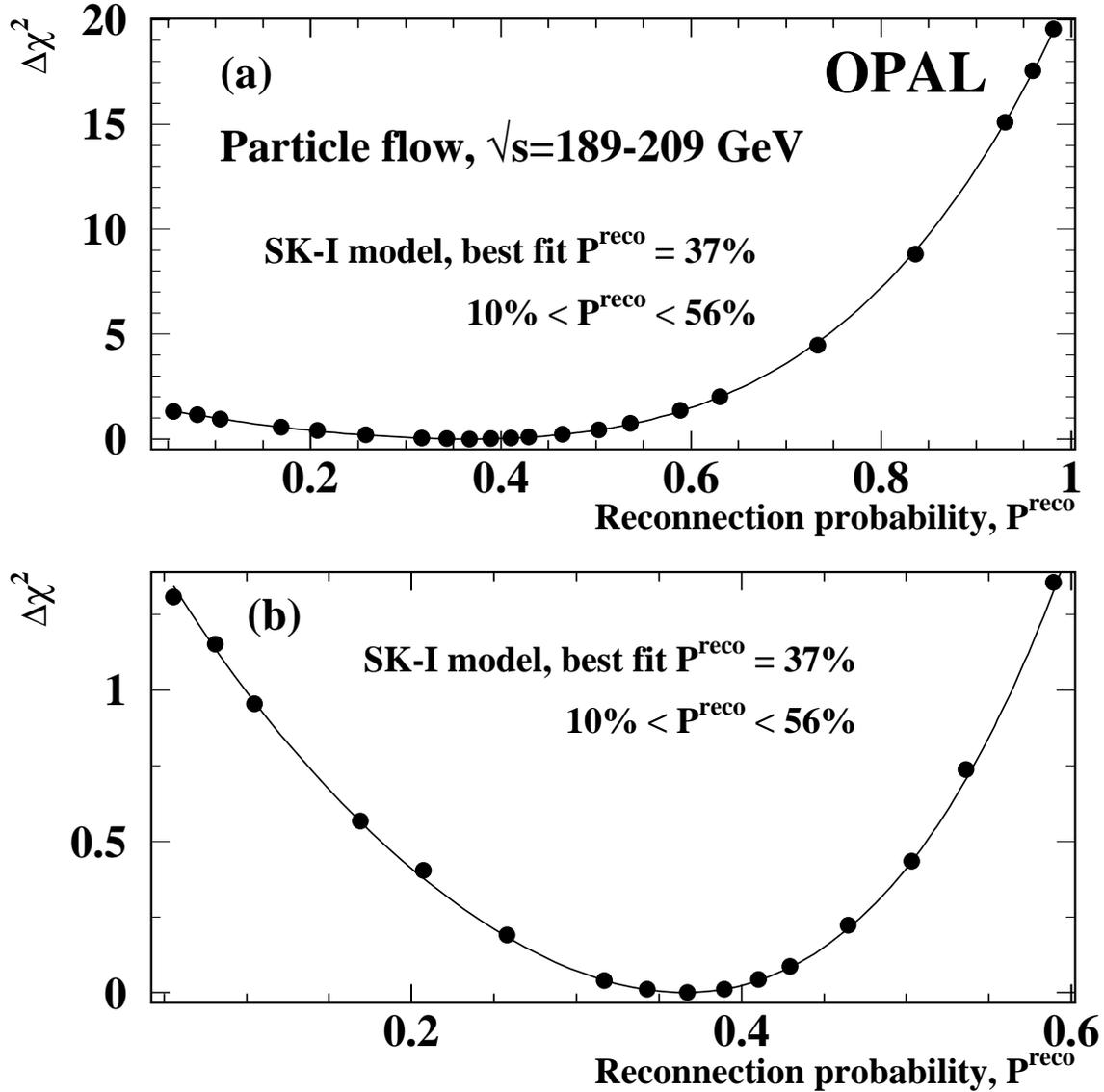,width=\textwidth}}
 \caption{$\Delta\chi^2$ curve obtained from comparison of the average
 \Rn\ measured using OPAL data between 189~\GeV\ and 209~\GeV, with the
 predictions of the \SKI\ model as a function of the fraction of
 reconnected events, \Prec, carried out at a reference centre-of-mass
 energy of 199.5~\GeV.  (a) shows the entire range of \Prec, while (b)
 shows the lower \Prec\ range of (a) in more detail.  The best
 agreement between the model and data is obtained when 37\% of events
 are reconnected in this model.}
 \label{fig:ski_scan}
\end{figure}

 Comparing the measured \Rn\ with the predictions of
 Table~\ref{tab:results}, the data are seen to be closest to the
 predictions of the \SKI\ model with strength parameter $\kI=0.9$.  As
 this parameter is arbitrary, it may be varied to optimise the
 consistency with the measured \Rn\ of Equation~\ref{eq:rn_avge}.  The
 $\Delta\chi^2$ curve corresponding to this variation is presented in
 Figure~\ref{fig:ski_scan}. Parametrising this curve using a fourth
 order polynomial, the best agreement with data is obtained when
 approximately 37\% of events are reconnected in the \SKI\ model,
 corresponding to the value $\kI=1.0$.
  The 68\% confidence level allowed region deduced from the
 $\Delta\chi^2$ curve corresponds to $0.10< \Prec < 0.56$.  This
 result is not combined with the analysis of inclusive particle
 multiplicity as they have correlated systematic uncertainties and no
 significant improvement in sensitivity is expected.

  It should be noted that the properties of the \SKI\ model vary
 significantly with the parton shower cut-off parameter and therefore
 this \kI\ cannot be directly compared to similar results from other
 LEP collaborations.  Any combination of results from the different
 experiments is best performed on the basis of analysis of a common
 set of simulated events, analysed independently by each experimental
 collaboration \cite{bib:lepewwg}.

\section{Conclusions}
\label{sec:conclusions}

   The predictions of models of colour reconnection implemented within
   the \Ariadne\ Monte Carlo, the \Herwig\ model and the \SK\ model,
   have been compared with OPAL data recorded at
   $\roots\simeq$189--209~\GeV\ using both inclusive measurements of
   particle multiplicity and a generalisation of the ``string effect''
   analysis to the four-jet topology of \WWqqqq\ events.

  Studies of reconnection phenomena using the extreme scenarios of the
  \SKI\ model show that changes up to approximately 1\% may be
  expected in \nchQQQQ, where the total experimental uncertainty on
  measurements of \nchQQQQ\ is 0.7\%.  Other models predict somewhat
  smaller effects.  Defining \Dnch\ using data alone provides a
  model-independent (but less sensitive) test of possible reconnection
  effects.  The inclusive measurements of particle multiplicity find
  no evidence for such effects.

   Measurements of particle flow in the OPAL data exclude an extreme
   scenario of the \SKI\ model and are compatible with other CR models
   such as \SKII, \SKIIpr, \ARII\ and that of \Herwig. They are also
   compatible with models which do not include colour reconnection,
   slightly disfavouring the conventional \Herwig\ model.  The results
   of this analysis are not combined with measurements of inclusive
   particle multiplicity as they have correlated systematic
   uncertainties and no significant improvement in sensitivity is
   expected. The best agreement with data is obtained using the \SKI\
   model with a reconnection probability, \Prec, of approximately
   37\%, corresponding to a model parameter $\kI=1.0$ within the
   context of the \Opal\ tuning of the \Jetset\ hadronisation model.
   The 68\% confidence level allowed region deduced from the $\chi^2$
   curve corresponds to $0.10 < \Prec < 0.56$. This result is used to
   help constrain the systematic uncertainty related to colour
   reconnection in measurements of the W boson mass
   \cite{bib:opalmw_final}.

\medskip
\bigskip\bigskip
\appendix
\par
\section*{Acknowledgements}
We particularly wish to thank the SL Division for the efficient
operation of the LEP accelerator at all energies and for their close
cooperation with our experimental group.  In addition to the support
staff at our own institutions we are pleased to acknowledge the \\
Department of Energy, USA, \\ National Science Foundation, USA, \\
Particle Physics and Astronomy Research Council, UK, \\ Natural
Sciences and Engineering Research Council, Canada, \\ Israel Science
Foundation, administered by the Israel Academy of Science and
Humanities, \\ Benoziyo Center for High Energy Physics,\\ Japanese
Ministry of Education, Culture, Sports, Science and Technology (MEXT)
and a grant under the MEXT International Science Research Program,\\
Japanese Society for the Promotion of Science (JSPS),\\ German Israeli
Bi-national Science Foundation (GIF), \\ Bundesministerium f\"ur
Bildung und Forschung, Germany, \\ National Research Council of
Canada, \\ Hungarian Foundation for Scientific Research, OTKA
T-038240, and T-042864,\\ The NWO/NATO Fund for Scientific Research,
the Netherlands.\\

\clearpage


\begin{table}[htbp]
 \begin{center}
 \begin{tabular}{|l|c|c|c|c|} \hline
  Multiplicity        & \nchQQQQ & \nchQQLV & \Dnch & \nchW  \\ \hline
   Data                  & 38.74  &  19.39 & $-0.04$ &  19.38 \\
   Stat. error           &  0.12  &   0.11 &   0.25  &   0.05 \\ \hline
   Systematics           &        &        &         &        \\ \hline
   \WW\ hadronisation    &  0.22  &  0.08  &  0.08   & 0.06   \\
   BEC                   &  0.03  &  0.01  &  0.05   & 0.01   \\
   Track definition      &  0.09  &  0.03  &  0.09   & 0.04   \\
   \eeqq\ modelling      &  0.11  &  0.01  &  0.10   & 0.04   \\
   \eeqq\ rate           &  0.01  &  0.00  &  0.02   & 0.01   \\
   Four-fermion background modelling
                         &  0.01  &  0.02  &  0.02   & 0.01   \\
   \ZZ\ rate             &  0.01  &  0.01  &  0.02   & 0.00   \\
   Residual backgrounds  &  0.00  &  0.01  &  0.02   & 0.00   \\
   Unfolding procedure   &  0.00  &  0.01  &  0.02   & 0.00   \\

\hline                                                    
 Total syst.             &  0.26  &  0.09  &  0.17  & 0.08   \\ \hline
 \end{tabular}
 \end{center}
\caption{Results and estimated systematic effects in inclusive
         charged particle multiplicity measurements, see text for
         details.}
\label{tab:systematics_nch}
\end{table}

\begin{table}[tbph]
 \begin{center}
 \begin{tabular}{|l|c|c|c|c|} \hline
Sample            & \Dnch   & significance(\Dnch) 
                  & \Rn     & significance(\Rn)           \\ \hline

Data              & $-0.04\pm0.30$ &  & $1.243\pm0.034$&   \\\hline
\SKI  ($\kI=100$) & $-0.42$   & $+1.2$& 1.092   & $+4.4$    \\
\SKI  ($\kI=0.9$) & $-0.29$   & $+0.8$& 1.246   & $-0.1$   \\
\SKII             & $-0.14$   & $+0.3$& 1.273   & $-0.9$   \\
\SKIIpr           & $-0.16$   & $+0.4$& 1.277   & $-1.0$   \\
\ARII             & $-0.19$   & $+0.5$& 1.271   & $-0.8$   \\
\Herwig-CR        & $+0.32$   & $-1.2$& 1.282   & $-1.2$   \\ \hline
\Jetset\          & $-0.04$   & $ 0.0$& 1.291   & $-1.4$   \\
\Herwig\          & $+0.02$   & $-0.2$& 1.311   & $-2.0$   \\
\Ariadne\         & $+0.02$   & $-0.2$& 1.286   & $-1.3$   \\
Old \Jetset\        
                  & $-0.03$   & $ 0.0$& 1.280   & $-1.1$   \\
\ARI\             & $ 0.00$   & $-0.2$& 1.304   & $-1.8$   \\ \hline
 \end{tabular}
 \end{center}
\caption{Comparison of the average measured \Dnch\ and \Rn\ in data
 with various models at $\roots=199.5$~\GeV.  The level of agreement
 in \Rn\ is given by the significance(\Rn), \ie,
 $(\Rn(\mathrm{data})-\Rn(\mathrm{model}))/\sigma^{\mathrm{total}}_{\Rn}$,
 the difference between the average value of \Rn\ in data and each
 model divided by the total uncertainty, and similarly for \Dnch.}
\label{tab:results}
\end{table}

\begin{table}[tbph]
 \begin{center}
 \begin{tabular}{|l|c|c|c|c|c|} \hline
$\langle \roots \rangle$ (\GeV) & $\int{\cal L}dt$ (pb$^{-1}$) & Selected events &
       Efficiency (\%)   & Purity (\%) & Correct jet pairing (\%) \\ \hline
 188.6       & 183.0& 675     & 39.7     & 86.1        & 90.3 \\
 191.6       & 29.3 &  92     & 39.1     & 86.5        & 90.0 \\
 195.5       & 76.4 & 277     & 40.0     & 87.1        & 89.9 \\
 199.5       & 76.6 & 253     & 38.9     & 87.0        & 89.4 \\
 201.6       & 37.7 & 145     & 38.4     & 84.2        & 89.1 \\
 206.0       & 220.5& 757     & 37.7     & 86.2        & 88.5 \\
\hline
 \end{tabular}
 \end{center}
\caption{Summary of the integrated luminosity and number of candidate
events used in the particle flow analysis, after all selection
criteria, at each centre-of-mass energy in the range 189-209~\GeV.  The
efficiency and purity are defined relative to the \WWqqqq\ production
process.  The ``correct'' pairing in the rightmost column is defined
by whichever association of observed jets in the detector minimises
the sum of angular differences relative to the original four fermions
from the \WW\ decay.}
\label{tab:data_summary}
\end{table}

\begin{table}[tbph]
 \begin{center}
 \begin{tabular}{|l|c|c|} \hline
Model             & \Prec\ (\%) & \sensRn  \\ \hline
\SKI  ($\kI=100$) & 98.2        & 7.9   \\
\SKI  ($\kI=0.9$) & 34.3        & 1.7   \\
\SKII             & 17.2        & 0.6   \\
\SKIIpr           & 16.1        & 0.5   \\
\ARII             & 49.4        & 1.3   \\
\Herwig-CR        & 23.0        & 0.9   \\ \hline
 \end{tabular}
 \end{center}
\caption{Summary of the predicted statistical sensitivity of the
  particle flow analysis for different models of colour reconnection
  at $\roots=199.5$~\GeV.  The sensitivity is defined as the
  difference between a given reconnection model and its corresponding
  ``no reconnection'' sample ($\Delta\Rn$), divided by the expected
  error of all data combined (\sigRnstat).  \Prec\ is the fraction of
  colour reconnected events in each model.  For \ARII, the no-CR model
  is \ARI.}
\label{tab:sensitivity}
\end{table}

 \begin{table}[tbph]
  \begin{center}
  \begin{tabular}{|l|c|} \hline
  Particle Flow        & \Rn    \\ \hline
   Data                & 1.243  \\
   Stat.\ error        & 0.025  \\ \hline
   Systematics         &        \\ \hline
  \WW\ hadronisation   &0.015    \\
  BEC                  &0.002    \\
  Track definition     &0.014    \\
  \eeqq\ modelling     &0.010   \\ 
  \eeqq\ rate          &0.002    \\
  Four-fermion background modelling  &0.002    \\
  \ZZ\ rate            &0.001    \\
  Residual backgrounds &0.000    \\ \hline
 Total syst.           &0.023 \\ \hline
  \end{tabular}
  \end{center}
 \caption{Result and estimated systematic effects in particle
 flow measurements, see text for details.}
 \label{tab:systematics}
 \end{table}

\end{document}